\documentclass[12pt]{article}  

\usepackage{graphicx,color,epsf} 

\usepackage{amsmath,latexsym}

\newcommand{\lsim}{\raisebox{-0.13cm}{~\shortstack{$<$ \\[-0.07cm]
      $\sim$}}~}
\begin{document}

\begin{center}
{\LARGE \bf Ambiguities  in the 
local thermal behavior
of the scalar radiation 
in one-dimensional boxes}
\\ 
\vspace{1cm} 
{\large E. S. Moreira Jr.}
\footnote{E-mail: moreira@unifei.edu.br}   
\\ 
\vspace{0.3cm} 
{\em Instituto de Matem\'{a}tica e Computa\c{c}\~{a}o,}  
{\em Universidade Federal de Itajub\'{a},}   \\
{\em Itajub\'a, Minas Gerais 37500-903, Brazil}

\vspace{0.3cm}
{\large May, 2020}
\end{center}
\vspace{0.5cm}


\abstract{ 
This paper reports certain ambiguities in the calculation of the 
ensemble average 
$\left<T_\mu{}_\nu\right>$ 
of the stress-energy-momentum tensor of an arbitrarily coupled massless scalar field
in one-dimensional boxes in flat spacetime. The study addresses a
box with periodic boundary condition (a circle) 
and boxes with reflecting edges (with Dirichlet's or Neumann's  boundary conditions at the endpoints).
The expressions for  $\left<T^\mu{}^\nu\right>$  are obtained from
finite-temperature Green functions. 
In an appendix, in order to control divergences typical of two dimensions, 
these Green functions are calculated   
for related backgrounds with arbitrary number of  dimensions and for scalar fields of arbitrary mass, and specialized in the text to two dimensions and for massless fields.
The ambiguities arise due to the presence in 
$\left<T^\mu{}^\nu\right>$
of double series that are not 
absolutely convergent. 
The order in which the two associated summations are evaluated matters, leading to two different thermodynamics for each type of box. In the case of a circle, 
it is shown that the ambiguity corresponds to the classic controversy in the literature whether or not zero mode contributions should be taken into account in computations of partition functions. 
In the case of boxes with reflecting edges, it results that one of the 
thermodynamics corresponds to
a total energy 
(obtained by integrating the non homogeneous energy density over space)
that does not depend on the curvature coupling parameter $\xi$ as expected;
whereas the other thermodynamics curiously corresponds
to a total energy that does depend on $\xi$.
Thermodynamic requirements (such as local and global stability) and their restrictions to the values of $\xi$ are considered. 
}

\vspace{0.5cm}
PACS number(s): 11.10.Wx, 04.62.+v, 05.70.-a

\section{Introduction}
\label{introduction}
Over the last decades, since the discovery that a black hole behaves
very much like  a  blackbody, with entropy and temperature \cite{bek73}, 
and even radiating \cite{haw75},
the study of quantum fields at finite temperature near boundaries
and in spacetimes with non trivial topologies and geometries 
has received increasing attention in the literature.
The message seems to be that one may learn a great deal about the nature of gravity
itself by looking at boundary quantum field theory at finite temperature, 
especially in lower dimensions as holography suggests \cite{sus05}.


A simple example that is commonly used to
illustrate the interplay between thermodynamics of fields and non trivial topology is the model of a massless scalar field 
living on a circle of length $a$
and geometry (throughout the text $k_{B}=\hbar=c=1$),
\begin{equation}
ds^{2}=dt^{2}-dx^{2}.
\label{geometry}
\end{equation}
Familiar methods in statistical mechanics lead to the internal energy,
\begin{equation}
U(T,a)=-\frac{\pi}{6a}+\frac{4\pi}{a}\sum_{k=1}^{\infty}\frac{k}{e^{2\pi k/Ta}-1},
\label{ie-circle}
\end{equation} 
where the first term is the vacuum energy (i.e., corresponding to 
$T\rightarrow 0$) \cite{dav82}, and the second is the contribution
at temperature  $T$ due to the Planck distribution \cite{hua87}.
In fact eq. (\ref{ie-circle}) disguises a subtlety that has simply been ignored.
It turns out that due to the periodic boundary condition proper of a circle, in evaluating $U$ one should also take into account the mode corresponding
to $k=0$, i.e., the so called ``zero mode'' \cite{dow88}.
However, in order to do so
some regularization must be used. For example, the  term,
\footnote{Note the factor $2\pi/a$ in eq. (\ref{regularization}), and not $4\pi/a$
as in eq. (\ref{ie-circle}) where each term in the sum\-ma\-tion cor\-re\-sponds to two states.
}
\begin{equation}
\frac{2\pi}{a}\lim_{\epsilon\rightarrow 0}
\frac{\epsilon}{e^{2\pi \epsilon/Ta}-1},
\label{regularization}
\end{equation}
could be added to eq. (\ref{ie-circle}) resulting in,
\begin{equation}
U(T,a)=-\frac{\pi}{6a}+T+\frac{4\pi}{a}\sum_{k=1}^{\infty}\frac{k}{e^{2\pi k/Ta}-1}.
\label{ie-zeromode}
\end{equation} 
The contrasting expressions 
in eqs. (\ref{ie-circle}) and (\ref{ie-zeromode}) are source of
a dispute in the literature regarding whether the zero mode should be taken into
account or not \cite{bre02,dow03,eli02}. This issue is relevant since it is related
with the third law of thermodynamics \cite{dow88,bre02,eli02}, 
the derivation of the Cardy-Verlinde formula and
entropy bounds \cite{kut01,kle01}, among other topics \cite{yaz17}.

Arguing in favour of eq. (\ref{ie-circle}) the authors of ref. \cite{bre02}
remark that an independent calculation by using the thermal Green function
leads to a homogeneous energy density $\left<T_{t}{}_{t}\right>$ which multiplied
by the length $a$ yields precisely $U$ in eq. (\ref{ie-circle}). Indeed, when
looking through the literature one finds an earlier calculation in a textbook, ref.
\cite{dav82}, where the ensemble average 
$\left<T_\mu{}_\nu\right>$
of the stress-energy-momentum tensor 
is determined from the thermal Green function. Then,
by taking  $\left<T_{t}{}_{t}\right> \times a$,
eq. (\ref{ie-circle}) comes up again. It is rather puzzling that 
such a ``local approach'' to obtain the internal energy ignores 
the term $T$ in eq. (\ref{ie-zeromode}).

The apparent absence of eq. (\ref{ie-zeromode}) in the ``local approach''
has motivated the investigation in section \ref{circle}, 
whose content is now outlined.
In appendix \ref{acircle}, the finite-temperature Green function for a 
neutral scalar field of mass $M$ in a flat $N$-dimensional spacetime
with periodic boundary condition along one of the dimensions
is calculated. By taking $M\rightarrow 0$ and $N\rightarrow 2$, the Green function
is used in section \ref{lcircle} to obtain $\left<T_\mu{}_\nu\right>$ of a massless scalar field on a circle of length $a$.
As is typical in this kind of calculation involving finite-temperature Green functions
to obtaining the ensemble average of the stress-energy-momentum tensor      \cite{bro69,dow77,alt78,tad86},
the resulting expression of the homogeneous $\left<T_\mu{}_\nu\right>$ contains a double series. 
In higher dimensions the double series converges absolutely
resulting that one can interchange the order of the summations and the result still
comes out the same (see, e.g., ref. \cite{rud76}). However, that is not the case in two-dimensional background
as the calculations in section  \ref{circle} show:
one order in which the summations are evaluated
leads indeed to eq. (\ref{ie-circle}); but interchanging the order of the summations
leads to eq. (\ref{ie-zeromode}) instead.

In section \ref{cstability}, local thermodynamic stability of the two sides of the ambiguity is addressed.
In section \ref{gcircle}, in order to explore further the implications of this ambiguity in thermodynamics, the formula 
(with $\beta:=1/T$), 
\begin{equation}
U=\left(\frac{\partial(\beta F)}{\partial \beta}\right)_{a},
\label{fu}
\end{equation}
is integrated to determine  the 
Helmholtz free energy $F$. 
The integration constant resulting from this procedure is set by requiring
that the thermodynamic pressure matches the stress component of 
$\left<T_\mu{}_\nu\right>$, such that there is no unknown length scale.
Various thermodynamics aspects are investigated at the asymptotic limits
when $Ta\ll 1$, and when $Ta\gg 1$.

It is rather well known that an interval with Neumann boundary condition at
the endpoints 
is a model of a one-dimensional
box with reflecting edges in which the calculation of the partition function
for hot scalar radiation also leads to a zero mode commonly ignored. 
In fact, the formula given in the literature for the internal energy corresponding to  Dirichlet's boundary condition at the endpoints (for which there is no zero mode)
is the same as that corresponding to Neumann's, namely \cite{amb83,lim09}, 
\begin{equation}
U(T,a)=-\frac{\pi}{24 a}+\frac{\pi}{a}\sum_{k=1}^{\infty}\frac{k}{e^{\pi k/Ta}-1},
\label{ie-rwall}
\end{equation} 
where $a$ is the length of the interval.
Considering this fact and the discussion in the previous paragraphs one may wonder if 
the local approach using Green functions may contain surprises here as well. In
examining the literature, this author has not found any study
of $\left<T_\mu{}_\nu\right>$ for the  hot scalar radiation with
Dirichlet's or Neumann's boundary conditions at the endpoints of a one-dimensional box
\footnote{In fact, there is such a calculation in four dimensions \cite{tad86}; 
but then there is no ambiguity since the corresponding double series are absolutely convergent.}. 
Such a study is implemented in section \ref{lwalls}, and outlined below.

In appendix \ref{arwalls}, 
the finite-temperature Green function for a 
neutral scalar field of mass $M$ in $N$-dimensional flat spacetime
with two parallel plane walls at which either Dirichlet's
or Neumann's boundary conditions are taken is evaluated.
Then, in section \ref{lwalls}, once more one sets
$M\rightarrow 0$ and $N\rightarrow 2$ and uses the Green function to obtain 
$\left<T_\mu{}_\nu\right>$ for a massless scalar field in a one-dimensional box 
of length $a$
with reflecting endpoints. This time, it turns out that $\left<T_\mu{}_\nu\right>$ 
is non homogeneous and dependent on the curvature coupling parameter $\xi$.
Ambiguities now arise due to the presence in
$\left<T_\mu{}_\nu\right>$ 
of two sets of double series. A particular order of summation is chosen in each series,
and then the order is interchanged, resulting in two different 
expressions for $\left<T_\mu{}_\nu\right>$. The corresponding
local thermal behaviors are investigated near the endpoints and at the midpoint of the box, for low and high temperatures. The values of $\xi$ consistent with local stable thermodynamic equilibrium (see ref. \cite{del15}) are determined in section \ref{istability}.

In section \ref{gwalls}, in order to calculate the internal energies corresponding
to the two expressions for $\left<T_\mu{}_\nu\right>$ found in section \ref{lwalls}, one integrates the non homogeneous energy densities $\left<T_{t}{}_{t}\right>$ over the box. It is then shown that one of the integrations leads to the formula in the literature (calculated using the partition function), i.e.,
eq. (\ref{ie-rwall}), whereas the other integration yields instead,
\begin{equation}
U(T,a)=-\frac{\pi}{24 a}+
\left(1\mp 4\xi\right)\frac{T}{2}+
\frac{\pi}{a}\sum_{k=1}^{\infty}\frac{k}{e^{\pi k/Ta}-1},
\label{ie-rwall-zeromode}
\end{equation} 
where, as in the rest of the text, the upper sign applies to Dirichlet's boundary condition and the lower sign to
Neumann's. It is worth noting that eqs. (\ref{ie-circle}) and (\ref{ie-zeromode})  correspond to eqs. (\ref{ie-rwall}) and (\ref{ie-rwall-zeromode}), respectively.

The appearance of $\xi$ in eq. (\ref{ie-rwall-zeromode}) is a bit surprising since for
a massless scalar field $\phi$ in flat spacetime with one spatial dimension $x$,
the classical expression for the energy density $T_{tt}$ depends on the
curvature coupling parameter $\xi$ only through the term 
$-2\xi\partial_{x}(\phi\partial_{x}\phi)$, 
which thus does not contribute when integrating $T_{tt}$
for Dirichlet or Neumann boundary conditions \cite{ful07}.
It should also be noticed that by setting $\xi=1/4$ in eq. (\ref{ie-rwall-zeromode}),
Neumann's zero mode mentioned above emerges. A last inescapable remark on 
eq. (\ref{ie-rwall-zeromode}) at this early
stage in the paper regards the low temperature behavior of the corresponding heat capacity at constant volume, namely 
$C=(1\mp 4\xi)/2$ (up to a positive exponential small correction). As thermal stability requires
$C> 0$ \cite{cal85}, it follows that $\xi$ must be such that $\xi\leq 1/4$ for Dirichlet's, 
and $\xi\geq -1/4$ for Neumann's. Later on in the text, these inequalities will be confronted
with those obtained in section \ref{istability} 
where local stable thermodynamic equilibrium is required \cite{del15}.

The rest of the material in section \ref{gwalls} investigates further
the two thermodynamics corresponding to eqs. (\ref{ie-rwall}) and (\ref{ie-rwall-zeromode}).
Again eq. (\ref{fu}) is used to obtain $F$, and from that the other thermodynamic quantities,
whose behaviors are studied when $Ta\ll 1$, and when $Ta\gg 1$.
As in section \ref{gcircle}, various thermodynamic aspects are addressed.
It should be mentioned that, unlike the other cases, by requiring that the thermodynamic pressure
be equal to the stress component of 
$\left<T_\mu{}_\nu\right>$  corresponding to eq. (\ref{ie-rwall-zeromode}), this time a  
length scale arises.

Section \ref{conclusion} contains a summary and further discussion
on the results.



\section{Circle}
\label{circle}
In this section the thermal behavior of 
a massless scalar field living on a circle
of length $a$ will be considered. Therefore
the geometry is that in eq. (\ref{geometry}) and
the endpoints $x=0$ and $x=a$ are identified.

\subsection{$\left<T_\mu{}_\nu\right>$ }
\label{lcircle}
The ensemble average $\left<T_\mu{}_\nu\right>$ can be formally obtained by using the ``point splitting'' method 
to the Feynman propagator $G_{{\cal F}}({\rm x},{\rm x}')$ at finite
temperature $T=1/\beta$
(see, e.g.,  refs. \cite{dav82,ful89} or the short review in ref. \cite{mor19}). In a flat background,
$$
\langle T_{\mu\nu}\rangle=
i\lim_{{\rm x'}\rightarrow {\rm x}}
[(1-2\xi)\nabla_{\mu}\nabla_{\nu '}
+(2\xi-1/2)g_{\mu\nu}
\nabla_{\sigma}\nabla^{\sigma '}
-2\xi\nabla_{\mu}\nabla_{\nu }]
G_{{\cal F}}({\rm x},{\rm x}'),
$$
where 
$G_{{\cal F}}({\rm x},{\rm x}')$
is related to 
the Green functions calculated in
appendix \ref{green-functions} by,
\begin{equation}
G_{E}({\rm x},{\rm x}')=iG_{{\cal F}}({\rm x},{\rm x}').
\label{feynman}
\end{equation}
For the analysis that will be implemented ahead, it is convenient to express each component of $\left<T_\mu{}_\nu\right>$ explicitly.
It follows that the energy density is given by,
\begin{equation}
\left<T_{t}{}_{t}\right> =\frac{i}{2}\lim_{{\rm x}'\rightarrow{\rm x}}
\left[\frac{\partial^{2}}{\partial t\,\partial t'}
+(1-4\xi)\frac{\partial^{2}}{\partial x\,\partial x'}-4\xi\frac{\partial^{2}}{\partial t^{2}}
\right]G_{{\cal F}}({\rm x},{\rm x}'),
\label{edensity}
\end{equation}
the stress by,
\begin{equation}
\left<T_{x}{}_{x}\right> =\frac{i}{2}\lim_{{\rm x}'\rightarrow{\rm x}}
\left[\frac{\partial^{2}}{\partial x\,\partial x'}
+(1-4\xi)\frac{\partial^{2}}{\partial t\,\partial t'}-4\xi\frac{\partial^{2}}{\partial x^{2}}
\right]G_{{\cal F}}({\rm x},{\rm x}'),
\label{stress}
\end{equation}
and the fluxes by,
\begin{eqnarray}
&&
\left<T_{t}{}_{x}\right> =i\lim_{{\rm x}'\rightarrow{\rm x}}
\left[(1-2\xi)\frac{\partial^{2}}{\partial t\,\partial x'}
-2\xi\frac{\partial^{2}}{\partial t\,\partial x}
\right]G_{{\cal F}}({\rm x},{\rm x}'),
\label{efdensity}
\end{eqnarray}
and,
\begin{eqnarray}
&&
\left<T_{x}{}_{t}\right> =i\lim_{{\rm x}'\rightarrow{\rm x}}
\left[(1-2\xi)\frac{\partial^{2}}{\partial x\,\partial t'}
-2\xi\frac{\partial^{2}}{\partial x\,\partial t}
\right]G_{{\cal F}}({\rm x},{\rm x}').
\label{mfdensity}
\end{eqnarray}
The reason for the appearance of the curvature coupling parameter $\xi$
in eqs. (\ref{edensity}) to (\ref{mfdensity}), in spite of the flat geometry in eq. (\ref{geometry}),
is due to the fact that $T_\mu{}_\nu$ is defined by functional derivative with respect
to an arbitrary metric \cite{dav82}.

Considering eqs. (\ref{feynman}), (\ref{green1}) and (\ref{cvariable}), one has that
(keeping for the time being $N$ and $M$ arbitrary such that divergences can be properly controlled),
\begin{equation}
G_{{\cal F}}({\rm x},{\rm x}')=
\sum_{m=-\infty}^{\infty}
\sum_{n=-\infty}^{\infty}
f\left(\sigma^{(m,n)}\right),
\label{feynman2}
\end{equation}
where,
\begin{equation}
f(u):=
-\frac{i}{(2\pi)^{N/2}}M^{\frac{N-2}{2}}
(-u)^{\frac{2-N}{4}}
K_{\frac{N-2}{2}}\left(M\sqrt{-u}\right).
\label{f}
\end{equation}
It is convenient to break up the expression in eq. (\ref{feynman2}) into the following contributions,
\begin{eqnarray}
&&
G_{0}({\rm x},{\rm x}'):=f\left(\sigma^{(0,0)}\right),
\hspace{2.2cm}
G_{{\tt vacuum}}({\rm x},{\rm x}'):=\sideset{}{'}\sum_{n=-\infty}^{\infty}
f\left(\sigma^{(0,n)}\right),
\label{feynman3}
\\
&&
G_{{\tt thermal}}({\rm x},{\rm x}'):=\sideset{}{'}\sum_{m=-\infty}^{\infty}
f\left(\sigma^{(m,0)}\right),\hspace{0.4cm}
G_{{\tt mixed}}({\rm x},{\rm x}'):=\sideset{}{'}\sum_{m=-\infty}^{\infty}
\sideset{}{'}\sum_{n=-\infty}^{\infty}
f\left(\sigma^{(m,n)}\right),
\label{feynman4}
\end{eqnarray}
with the prime in the summation indicating that the term corresponding to $m=0$ or $n=0$
should be excluded. Noting eqs. (\ref{feynman2}) and (\ref{f}), and considering the asymptotic behavior
of $K_{\nu}({\rm z})$ \cite{arf85}, one sees that 
$G_{0}({\rm x},{\rm x}')$ is the familiar vacuum propagator in Minkowski spacetime, 
$G_{{\tt vacuum}}({\rm x},{\rm x}')$ is the vacuum propagator due to a finite length $a$
(which vanishes if $a\rightarrow\infty$),
$G_{{\tt thermal}}({\rm x},{\rm x}')$ is the familiar thermal propagator in Minkowski spacetime  (corresponding to Planck's distribution, and thus vanishing if $T\rightarrow 0$), and that
$G_{{\tt mixed}}({\rm x},{\rm x}')$ has a ``mixed'' nature 
(vanishing if $a\rightarrow\infty$ or if $T\rightarrow 0$).
Since the background is flat, renormalization is implemented by dropping 
$G_{0}({\rm x},{\rm x}')$, yielding the renormalized propagator $G({\rm x},{\rm x}')$ that will replace
$G_{{\cal F}}({\rm x},{\rm x}')$
in  eqs. (\ref{edensity}) to (\ref{mfdensity}),
\begin{equation}
G({\rm x},{\rm x}')=
G_{{\tt vacuum}}({\rm x},{\rm x}')+G_{{\tt mixed}}({\rm x},{\rm x}')+G_{{\tt thermal}}({\rm x},{\rm x}').
\label{rfeynman}
\end{equation}

Setting now $N=2$ in eq. (\ref{f}), when $M\rightarrow 0$, it follows that \cite{arf85},
\begin{equation}
f(u)=\frac{i}{2\pi}\left[\ln\left(\frac{M\sqrt{-u}}{2}\right)+\gamma\right]+\cdots,
\hspace{1.0cm} N=2.
\label{fzero}
\end{equation}
Differentiating eq. (\ref{fzero}) and then setting $M=0$, it results that,
\begin{equation}
f'(u)=\frac{i}{4\pi u},
\label{fprime}
\end{equation}
which is used 
in  eqs. (\ref{edensity}) to (\ref{mfdensity}) to determine the components of 
$\left<T_\mu{}_\nu\right>$ 
when $N=2$ and $M=0$. (Clearly, for $N=2$, only the two first terms
in eq. (\ref{cvariable}) are taken into account.)
The action of the differential operators in  eqs. (\ref{edensity}) to (\ref{mfdensity})
on eq. (\ref{rfeynman}) yields,
\begin{equation}
\left<T_\mu{}_\nu\right>=
\left<T_\mu{}_\nu\right>_{{\tt vacuum}}+
\left<T_\mu{}_\nu\right>_{{\tt mixed}}+
\left<T_\mu{}_\nu\right>_{{\tt thermal}}.
\label{stress-tensor}
\end{equation}
Each contribution in eq. (\ref{stress-tensor}) is diagonal
with the energy densities equalling the corresponding stresses, and leading to,
\begin{equation}
\left<T_{x}{}_{x}\right>=\left<T_{t}{}_{t}\right>, 
\hspace{1.0cm}
\left<T_{t}{}_{x}\right>=\left<T_{x}{}_{t}\right>=0.
\label{components}
\end{equation}
According to eq. (\ref{edensity}),
\begin{equation}
\left<T_{t}{}_{t}\right>=
\left<T_{t}{}_{t}\right>_{{\tt vacuum}}+
\left<T_{t}{}_{t}\right>_{{\tt mixed}}+
\left<T_{t}{}_{t}\right>_{{\tt thermal}},
\label{ced}
\end{equation} 
where the first and the last terms are the well known expressions 
(two-dimensional versions)
for the ``Casimir'' energy density and the ``blackbody'' energy density, respectively,
\begin{equation}
\left<T_{t}{}_{t}\right>_{{\tt vacuum}}=-\frac{\pi}{6a^{2}},
\hspace{1.0cm}
\left<T_{t}{}_{t}\right>_{{\tt thermal}}=\frac{\pi}{6}T^{2}.
\label{cvted}
\end{equation}
The second term in eq. (\ref{ced}) arises when the differential operators in eq. (\ref{edensity})
act on $G_{{\tt mixed}}({\rm x},{\rm x}')$ in eq. (\ref{feynman4}), and therefore it  contains a double series.
When $N=4$ (which is the case in refs. \cite{bro69,tad86}), the order in which the corresponding summations are evaluated is irrelevant since the double series is absolutely convergent \cite{rud76}.
However, when $N=2$, the double series is not absolutely convergent and the order of evaluation of the
summations does matter, as will now be shown.

One may sum first over the ``boundary'' number $n$, corresponding to,
\begin{equation}
\left<T_{t}{}_{t}\right>_{{\tt mixed}}=T^{2}u(Ta),
\hspace{1.0cm}
u(q):=\frac{2}{\pi}\sum_{m=1}^{\infty}\sum_{n=1}^{\infty}
\frac{m^{2}-q^{2}n^{2}}{(m^{2}+q^{2}n^{2})^{2}}.
\label{u}
\end{equation}
Or, instead,  one may sum first over the ``thermal'' number $m$, i.e., 
\begin{equation}
\left<T_{t}{}_{t}\right>_{{\tt mixed}}=T^{2}v(Ta),
\hspace{1.0cm}
v(q):=\frac{2}{\pi}\sum_{n=1}^{\infty}\sum_{m=1}^{\infty}
\frac{m^{2}-q^{2}n^{2}}{(m^{2}+q^{2}n^{2})^{2}}.
\label{v}
\end{equation}
It can be noticed that,
\begin{equation}
v(q)=-\frac{1}{q^{2}}u(1/q).
\label{uv}
\end{equation}
In fact, using ref. \cite{pru86} or ref. \cite{mat17}, the first summations
in eqs. (\ref{u}) and (\ref{v}) can be evaluated to give, 
\begin{equation}
u(q)=-\frac{\pi}{6}+\frac{\pi}{q^{2}}\sum_{k=1}^{\infty}{\rm cosech}^{2}\left(\frac{k\pi}{q}\right),
\hspace{1.5cm}
v(q)=\frac{\pi}{6q^{2}}-\pi\sum_{k=1}^{\infty}{\rm cosech}^{2}\left(k \pi q\right).
\label{uandv}
\end{equation}
These expressions can be compared with each other by using the identity,
\begin{equation}
\sum_{k=1}^{\infty}\frac{k}{e^{2\pi qk}-1}=\frac{1}{4}\sum_{k=1}^{\infty}{\rm cosech}^{2}\left(k \pi q\right),
\label{planck}
\end{equation}
to recast ``Schl\"{o}milch's formula'' as, \footnote{``Schl\"{o}milch's formula''
is a Ramanujan type identity which has been used throughout the literature
in related contexts. See, e.g.,
eq. (40) in ref. \cite{dow02} and eq. (1) in ref. \cite{mo18}. 
[A typo has been detected in eq. (40) of ref. \cite{dow02}: the term $1/2\pi$ should be replaced
by $1/2$.]}
\begin{equation}
-\pi q
\sum_{k=1}^{\infty}{\rm cosech}^{2}\left(k\pi q\right)
-\frac{\pi}{q}
\sum_{k=1}^{\infty}{\rm cosech}^{2}\left(\frac{k\pi}{q}\right)
=1-\frac{\pi}{6}\left(q+\frac{1}{q}\right),
\label{ramanujan}
\end{equation}
where one can appreciate the symmetry $q\rightarrow 1/q$.
Considering then eq. (\ref{ramanujan}) in eq. (\ref{uandv}), it follows that,
\begin{equation}
v(q)=u(q)+\frac{1}{q},
\label{vu}
\end{equation}
showing that the double series in eqs. (\ref{u}) and (\ref{v}) are
indeed distinct. Now, noting eqs. (\ref{ced}), (\ref{cvted}) and (\ref{uandv}), $u(Ta)$ 
in  eq. (\ref{u})
leads to,
\begin{equation}
\left<T_{t}{}_{t}\right>=-\frac{\pi}{6a^{2}}+\frac{\pi}{a^{2}}
\sum_{k=1}^{\infty}{\rm cosech}^{2}\left(\frac{k\pi}{Ta}\right);
\label{birrel}
\end{equation}
whereas $v(Ta)$ 
in  eq. (\ref{v})
leads to,
\begin{equation}
\left<T_{t}{}_{t}\right>=-\frac{\pi}{6a^{2}}+
\frac{T}{a}+
\frac{\pi}{a^{2}}
\sum_{k=1}^{\infty}{\rm cosech}^{2}\left(\frac{k\pi}{Ta}\right),
\label{dowker}
\end{equation}
where eq. (\ref{vu}) has been used.
Thus, one ends up with two expressions for the energy density that differ from each other by 
a term linear in temperature: eq. (\ref{birrel}),
which is obtained by ``summing first over $n$'', and eq. (\ref{dowker}) which arises
by ``summing first over $m$''. In fact, by taking into account eq. (\ref{components}), it is seen that
for each order of summation chosen it corresponds a different $\left<T_\mu{}_\nu\right>$.
It should be noticed that eqs. (\ref{components}) and
(\ref{birrel}) are the result reported in
the literature \cite{dav82} (i.e., ``summing first over $n$'').

Noting eqs. (\ref{components}), (\ref{birrel}) and (\ref{dowker}), one  sees that
the terms carrying $\xi$ in eqs. (\ref{edensity}) to (\ref{mfdensity}) all canceled each other,
and that $\left<T_\mu{}_\nu\right>$ is traceless, as it should be. Also, since  $\left<T_\mu{}_\nu\right>$
is stationary and homogeneous, it is trivially conserved, i.e., 
$\left<T^\mu{}^\nu\right>_{\hspace{-0.05cm},\nu}=0$.

The physics of the ambiguity in eqs. (\ref{birrel}) and (\ref{dowker}),
together with eq. (\ref{components}), can be better explored 
by considering the asymptotic behaviors of $\left<T_\mu{}_\nu\right>$. This is done next.
\subsubsection{Summing first over n}
\label{nlcircle}

Using eq. (\ref{planck}) in the expression for $u$ in eq. (\ref{uandv}), some manipulation leads to,
\begin{equation}
u(q\rightarrow 0)=-\frac{\pi}{6}+\frac{4\pi}{q^{2}}e^{-2\pi/q},
\label{usq}
\end{equation}
where smaller exponential corrections have been omitted (as will always be omitted in the rest of the text). 
Then, at low temperatures or for small circles, eqs. (\ref{ced}) to (\ref{u}) with eq. (\ref{usq}) yield, 
\begin{equation}
\left<T_{t}{}_{t}\right>=-\frac{\pi}{6a^{2}}+\frac{4\pi}{a^{2}}e^{-2\pi/Ta},
\hspace{1.0cm} 
Ta\ll 1,
\label{nedsq}
\end{equation}
showing that the correction to the ``Casimir'' energy density  [see eq. (\ref{cvted})]
decreases exponentially when $Ta\rightarrow 0$.

Considering now eqs. (\ref{uv}) and (\ref{vu}), it results,
\begin{equation}
u(q)=-\frac{1}{q}-\frac{1}{q^{2}}u(1/q),
\label{uu}
\end{equation}
which with eq. (\ref{usq}) gives,
\begin{equation}
u(q\rightarrow \infty)=-\frac{1}{q}+\frac{\pi}{6q^{2}}-4\pi e^{-2\pi q}.
\label{ubq}
\end{equation}
Using now eq. (\ref{ubq}) in eq. (\ref{u}), one obtains from eqs. (\ref{ced}) and (\ref{cvted})
that,
\begin{equation}
\left<T_{t}{}_{t}\right>=\frac{\pi}{6}T^{2}-\frac{T}{a}-4\pi T^{2}e^{-2\pi Ta},
\hspace{1.0cm} 
Ta\gg 1.
\label{nedbq}
\end{equation}
Thus the ``blackbody'' energy density  [see eq. (\ref{cvted})]
drops by $T/a$ at high temperatures or for big circles.

It should be remarked that the behavior of $u$ for large values of $q$ [see eq, (\ref{ubq})]
has been determined from its behavior for small values of $q$ [see eq, (\ref{usq})] through
eq. (\ref{uu}). Such a 
feature is typical of quantum fields at finite temperature in backgrounds with
boundaries and it has been long known in the literature \cite{bro69,rav89}.
It is also worth remarking that the ``blackbody'' like energy density in eq. (\ref{nedbq})
and the ``Casimir'' like energy density in eq. (\ref{nedsq}) correspond to different regimes 
(of temperature and size) of the very same phenomenon. 
\subsubsection{Summing first over m}
\label{mlcircle}

Noting eqs. (\ref{birrel}) and (\ref{dowker}) and the text just after them,
one sees that the easiest way of getting the asymptotic behaviors corresponding to 
``summing first over $m$'' is to add $T/a$ to eqs. (\ref{nedsq}) and (\ref{nedbq}), namely,
\begin{eqnarray}
\left<T_{t}{}_{t}\right>=-\frac{\pi}{6a^{2}}+\frac{T}{a}+\frac{4\pi}{a^{2}}e^{-2\pi/Ta},
&&
Ta\ll 1,
\label{medsq}
\end{eqnarray}
and,
\begin{eqnarray}
\left<T_{t}{}_{t}\right>=\frac{\pi}{6}T^{2}-4\pi T^{2}e^{-2\pi Ta},
\hspace{0.8cm} 
&&
Ta\gg 1.
\label{medbq}
\end{eqnarray}
Alternatively, one could work as above, 
using eq. (\ref{planck}) in the expression for $v$ in eq. (\ref{uandv}), obtaining,
\begin{equation}
v(q\rightarrow \infty)=\frac{\pi}{6q^{2}}-4\pi e^{-2\pi q}.
\label{vbq}
\end{equation}
Then eqs. (\ref{ced}), (\ref{cvted}) and (\ref{v}), with eq. (\ref{vbq}),
yield eq. (\ref{medbq}).
Now, from eqs. (\ref{uv}) and (\ref{vu}), it follows that,
\begin{equation}
v(q)=\frac{1}{q}-\frac{1}{q^{2}}v(1/q),
\label{vv}
\end{equation}
which combined with eq. (\ref{vbq}) gives,
\begin{equation}
v(q\rightarrow 0)=\frac{1}{q}-\frac{\pi}{6}+\frac{4\pi}{q^{2}}e^{-2\pi/q}.
\label{vsq}
\end{equation}
Considering again eqs. (\ref{ced}), (\ref{cvted}) and (\ref{v}), now with eq. (\ref{vsq}),
one ends up with eq. (\ref{medsq}).

In comparing eq. (\ref{nedsq}) with eq. (\ref{medsq}) and  eq. (\ref{nedbq}) with eq. (\ref{medbq}),
it is seen that the exponential small correction to the ``Casimir'' energy density has been
replaced by a linear one, and that the linear correction  to the ``blackbody''
energy density has been replaced by a exponential small correction. These modifications
will have radical consequences in thermodynamics, as will be shown shortly.

\subsection{Local thermodynamic stability}
\label{cstability}

It is natural to wonder whether the requirement of local
thermodynamic stability (see, e.g., section 4 in ref. \cite{mor17}) might resolved
the ambiguity in eqs. (\ref{components}),
(\ref{birrel}) and (\ref{dowker}). 
Consider a small segment of the circle, and assume that the temperature
$T_{in}$ inside the segment differs (due to a fluctuation) slightly from 
$T_{out}$, which is the temperature outside the segment.
Conservation of momentum dictates that the power (energy per unity of time)
radiated out of the segment is proportional to the differences of stresses inside and outside,
i.e.,
\begin{equation}
\Phi=\left<T_{x}{}_{x}\right>_{in}-\left<T_{x}{}_{x}\right>_{out},
\label{power}
\end{equation}
up to a positive overal factor \cite{mor17}.
Below, the regimes $Ta\ll 1$ and $Ta\gg 1$ will be investigated.

\subsubsection{Summing first over n}
\label{ncstability}
Using eqs. (\ref{components}) and (\ref{nedsq}) in eq. (\ref{power}), it results,
\begin{equation}
\Phi=\frac{4\pi}{a^{2}}\left(e^{-2\pi/aT_{in}}-e^{-2\pi/aT_{out}}\right),
\hspace{1.0cm} 
Ta\ll 1.
\label{cnpower}
\end{equation}
Say that $T_{in}>T_{out}$, i.e.,  $\Phi$ in eq. (\ref{cnpower}) is positive.
Taking the derivative with respect to temperature of the energy density in eq. (\ref{nedsq}),
it follows that,
\begin{equation}
\frac{\partial}{\partial T}\left<T_{t}{}_{t}\right>=
\frac{8\pi^{2}}{T^{2}a^{3}}e^{-2\pi/Ta}>0.
\label{lcnheat-capacity}
\end{equation}
Thus, as $\Phi>0$, energy will leave the segment. Due to conservation of energy 
(i.e., energy in the segment will decrease)
and noticing eq. (\ref{lcnheat-capacity}),
$T_{in}$ will drop with the thermodynamic equilibrium being restored, as expected.

Considering now eqs. (\ref{components}) and (\ref{nedbq}) in eq. (\ref{power}), 
the leading contribution is,
\begin{equation}
\Phi=\frac{\pi}{6}\left(T_{in}^{2}-T_{out}^{2}\right),
\hspace{1.0cm} 
Ta\gg 1.
\label{cnpower2}
\end{equation}
Saying that 
$T_{in}>T_{out}$, $\Phi$ in eq. (\ref{cnpower2}) is positive and energy will leave the segment.
As, from eq. (\ref{nedbq}), $\partial_{T}\left<T_{t}{}_{t}\right>>0$,
then conservation of energy determines that $T_{in}$ will drop, 
and thermodynamic equilibrium will be restored again.

These results show that ``summing first over $n$'' is consistent with 
local thermodynamic stability.

\subsubsection{Summing first over m}
\label{mcstability}
Considering eqs. (\ref{components}) and (\ref{medsq}),  eq. (\ref{power}) yields, 
\begin{equation}
\Phi=\frac{1}{a}\left(T_{in}-T_{out}\right),
\hspace{1.0cm} 
Ta\ll 1,
\label{cmpower}
\end{equation}
up to exponential small corrections. 
Taking into account the leading contribution in eq. (\ref{medsq}),
\begin{equation}
\frac{\partial}{\partial T}\left<T_{t}{}_{t}\right>=
\frac{1}{a}>0.
\label{lcmheat-capacity}
\end{equation}
By repeating the argument above, if 
$T_{in}>T_{out}$ in eq. (\ref{cmpower}), 
then $\Phi>0$ and energy leaves the segment. It follows then from
eq. (\ref{lcmheat-capacity}) that $T_{in}$ drops, i.e., thermodynamic equilibrium is recovered.

By using eqs. (\ref{components}) and (\ref{medbq}) in eq. (\ref{power}), 
one also ends up with eq. (\ref{cnpower2}). The same argument
just after eq. (\ref{cnpower2}) shows that here as well thermodynamic equilibrium will be restored.

Therefore,  ``summing first over $m$'' is also consistent with 
local thermodynamic stability.



\subsection{Thermodynamics}
\label{gcircle}
The first step to obtain thermodynamics in the ``local approach'' is
to integrate the energy density over the box, i.e.,
\begin{equation}
U=\int_{0}^{a}\left<T_{t}{}_{t}\right>dx,
\label{ienergy}
\end{equation}
yielding the internal energy $U$. By noticing the identity in eq. (\ref{planck})
and considering eq. (\ref{ienergy}),
the homogeneous $\left<T_{t}{}_{t}\right>$ in 
eqs. (\ref{birrel}) (``summing first over $n$'') and (\ref{dowker}) (``summing first over $m$'')
lead to the contrasting expressions for $U$ in eqs. (\ref{ie-circle})  and (\ref{ie-zeromode}), respectively, which as mentioned previously are source of the zero mode controversy in the ``global approach'' 
[see the text just after eq. (\ref{ie-zeromode})]. Thermodynamic aspects of this ambiguity are
also better appreciated by looking at the asymptotic behaviors.
\subsubsection{Summing first over n}
\label{ngcircle}
Corresponding to eq. (\ref{nedsq}) one has from eq. (\ref{ienergy}) that,
\begin{equation}
U(T,a)=-\frac{\pi}{6a}+\frac{4\pi}{a}e^{-2\pi/Ta},
\hspace{1.0cm} 
Ta\ll 1,
\label{nesq}
\end{equation}
in agreement with an early calculation using the partition function 
(i.e., the ``global approach'')
on the circle \cite{amb83}.
As the heat capacity at constant volume is positive, i.e., 
$C:=\partial_ {T}U>0$ [see eq. (\ref{lcnheat-capacity})], one of the criteria for
global thermodynamic stability is satisfied \cite{cal85}. 

Now, using eq. (\ref{nesq}) in eq. (\ref{fu}) and integrating, it results that,
\begin{equation}
F(T,a)=-\frac{\pi}{6a}-2Te^{-2\pi/Ta},
\hspace{1.0cm} 
Ta\ll 1,
\label{nfsq}
\end{equation}
where the integration constant has been set such that
the thermodynamic pressure ${\tt p}:=-\partial_ {a}F$ equals the stress $\left<T_{x}{}_{x}\right>$
[see eqs. (\ref{components}) and (\ref{nedsq})]. It should be noticed that, since 
$\left<T_\mu{}_\nu\right>$
is tracelles, the equation of state,
\begin{equation}
U={\tt p}a,
\label{estate}
\end{equation}
holds, and that 
the ``Casimir force'' $-\pi/6a^{2}$ 
(which tends to contract the circle)
is weakened by an exponential small ``thermal'' contribution
[see ${\tt p}$ in eq. (\ref{nedsq})]. The entropy $S=-\partial_ {T}F$
following from eq. (\ref{nfsq}) is given by,
\begin{equation}
S(T,a)=2\left(\frac{2\pi}{Ta}+1\right)e^{-2\pi/Ta},
\hspace{1.0cm} 
Ta\ll 1.
\label{nssq}
\end{equation}
Then, when $T\rightarrow 0$, $S$ in eq. (\ref{nssq}) vanishes, i.e., 
the third law of thermodynamics is satisfied.

Now, corresponding to the energy density in eq. (\ref{nedbq}) one has that,
\begin{equation}
U(T,a)=\frac{\pi}{6}aT^{2}-T-4\pi aT^{2}e^{-2\pi Ta},
\hspace{1.0cm} 
Ta\gg 1,
\label{nebq}
\end{equation}
which agrees with early calculations where the ``global approach''
has been used \cite{amb83,lim09}. 
It follows from eq. (\ref{nebq}) that
$C:=\partial_ {T}U>0$, which as mentioned previously is one of the criteria for
global thermodynamic stability \cite{cal85}. Following the same steps applied
in the regime $Ta\ll 1$ above, it results that [see eq. (\ref{fu})],
\begin{equation}
F(T,a)=-\frac{\pi}{6}aT^{2}+T \ln(Ta)-2Te^{-2\pi Ta},
\hspace{1.0cm} 
Ta\gg 1,
\label{nfbq}
\end{equation}
with eq. (\ref{estate}) still holding, 
i.e., ${\tt p}$ is given by eq. (\ref{nedbq}) where  $\pi T^{2}/6$ is the ``blackbody radiation force'' and
$-T/a$ is the ``thermal Casimir force''
(which, unlike the blackbody contribution, tends to contract the circle). 
The asymptotic behavior of the entropy associated with eq. (\ref{nfbq}) is,
\begin{equation}
S(T,a)=\frac{\pi}{3}aT-\left[\ln(Ta)+1\right]
-2(2\pi Ta-1)e^{-2\pi Ta},
\hspace{1.0cm} 
Ta\gg 1,
\label{nsbq}
\end{equation}
becoming the entropy of the ``blackbody radiation'',  $\pi a T/3$, as $Ta\rightarrow\infty$.

In both regimes above (i.e., $Ta\ll 1$ and $Ta\gg 1$) one can check that $\partial_ {a}{\tt p}>0$,
which, in fact,  violates one of the criteria for global thermodynamic stability 
\footnote{$C>0$ implies thermal stability; $\partial_{a}{\tt p}<0$ implies mechanical stability.} 
\cite{cal85}.

\subsubsection{Summing first over m}
\label{mgcircle}
Repeating the procedures above, $U$
corresponding to eq. (\ref{medsq}) is given by eq. (\ref{nesq}) after adding $T$,
following that, 
\begin{equation}
F(T,a)=-\frac{\pi}{6a}-T\ln(Ta)-2Te^{-2\pi/Ta},
\hspace{1.0cm} 
Ta\ll 1,
\label{mfsq}
\end{equation}
instead of eq. (\ref{nfsq}). The equation of state, eq. (\ref{estate}), holds, and
therefore the ``Casimir force'' is now weakened by a term linear in temperature
[see ${\tt p}$ in eq. (\ref{medsq})]. It follows from eq. (\ref{mfsq}) that,
\begin{equation}
S(T,a)=\ln(Ta)+1
+2\left(\frac{2\pi}{Ta}+1\right)e^{-2\pi/Ta},
\hspace{1.0cm} 
Ta\ll 1,
\label{mssq}
\end{equation}
which clearly violates the third law of thermodynamics, with the entropy $S$ 
diverging to $-\infty$ as $T\rightarrow 0$.
This fact is sometimes used in the literature to argue that the zero mode should not be taking into account
in computations of the partition function (see e.g. refs. \cite{bre02,eli02}), 
i.e., one should ``sum first over $n$'', accordingly.   

Turning now to the regime $Ta\gg 1$, 
corresponding to eq. (\ref{medbq}),  it follows that $U$ is given in eq. (\ref{nebq}) by omitting $-T$,
$F$ is given in eq. (\ref{nfbq}) by omitting $T\ln(Ta)$, and $S$
is given in eq. (\ref{nsbq}) by omitting $-\ln(Ta)-1$. The equation of state eq. (\ref{estate}) holds,
and ${\tt p}$ in eq. (\ref{medbq}) shows that there is no ``thermal Casimir force'', this time.

In both regimes it can be checked that $C>0$ and $\partial_{a}{\tt p}>0$, again.

\section{Interval with reflecting edges}
\label{interval}
This section addresses the thermal behavior of a massless scalar field in an interval 
where Dirichlet's or Neumann's  boundary conditions are taken at the endpoints, $x=0$ and $x=a$.
That is, the endpoints are the reflecting ``walls'' of a one-dimensional box 
containing hot scalar radiation
in flat two-dimensional spacetime [see eq. (\ref{geometry})].

\subsection{$\left<T_\mu{}_\nu\right>$}
\label{lwalls}

In order to determine $\left<T_\mu{}_\nu\right>$
\footnote{When $a\rightarrow\infty$ is set in $\left<T_\mu{}_\nu\right>$ obtained in this section,
formulas corresponding to the presence of a single reflecting wall at $x=0$ 
are consistently reproduced (see refs. \cite{del15,mor19,mor15}).}, one again uses 
eqs. (\ref{feynman}) to (\ref{mfdensity}), but now with the Feynman propagator given by
[see eq. (\ref{green2}),  eq. (\ref{dncvariable}), and text closing appendix \ref{arwalls}],
\begin{equation}
G_{{\cal F}}({\rm x},{\rm x}')=
\sum_{m=-\infty}^{\infty}
\sum_{n=-\infty}^{\infty}
\left[
f\left(\sigma_{-}^{(m,n)}\right)\mp f\left(\sigma_{+}^{(m,n)}\right)
\right],
\label{feynman5}
\end{equation}
with $f$ defined in eq. (\ref{f}) and recalling that the upper sign is for Dirichlet's
whereas the lower sign is for Neumann's boundary conditions. Removing from $G_{{\cal F}}({\rm x},{\rm x}')$
the Minkowski vacuum propagator $G_{0}({\rm x},{\rm x}')$ in eq. (\ref{feynman3})
[note that $\sigma^{(m,0)}=\sigma_{-}^{(m,0)}$], it results the following renormalized
propagator,
\begin{equation}
G({\rm x},{\rm x}')=
G_{{\tt vacuum}}^{Casimir}({\rm x},{\rm x}')
+
G_{{\tt vacuum}}^{wall}({\rm x},{\rm x}')
+G_{{\tt mixed}}^{Casimir}({\rm x},{\rm x}')
+G_{{\tt mixed}}^{wall}({\rm x},{\rm x}')
+G_{{\tt thermal}}({\rm x},{\rm x}'),
\label{dnrfeynman}
\end{equation}
where $G_{{\tt thermal}}({\rm x},{\rm x}')$ is the ``blackbody'' propagator in eq. (\ref{feynman4}) and, 
\begin{eqnarray}
&&
\hspace{-1.2cm}
G_{{\tt vacuum}}^{Casimir}({\rm x},{\rm x}'):=\sideset{}{'}\sum_{n=-\infty}^{\infty}
f\left(\sigma^{(0,n)}_{-}\right),
\hspace{1.6cm}
G_{{\tt vacuum}}^{wall}({\rm x},{\rm x}'):=\mp\sum_{n=-\infty}^{\infty}
f\left(\sigma^{(0,n)}_{+}\right),
\label{feynman6}
\\
&&
\hspace{-1.2cm}
G_{{\tt mixed}}^{Casimir}({\rm x},{\rm x}'):=\sideset{}{'}\sum_{m=-\infty}^{\infty}
\sideset{}{'}\sum_{n=-\infty}^{\infty}
f\left(\sigma^{(m,n)}_{-}\right),\hspace{0.5cm}
G_{{\tt mixed}}^{wall}({\rm x},{\rm x}'):=\mp \sideset{}{'}\sum_{m=-\infty}^{\infty}
\sum_{n=-\infty}^{\infty}
f\left(\sigma^{(m,n)}_{+}\right).
\label{feynman7}
\end{eqnarray}
One sees from these definitions and from eq. (\ref{f}) that, when $T\rightarrow 0$, only
the vacuum contributions remain in eq. (\ref{dnrfeynman}). When $a\rightarrow\infty$, both
$G_{{\tt vacuum}}^{Casimir}({\rm x},{\rm x}')$ and
$G_{{\tt mixed}}^{Casimir}({\rm x},{\rm x}')$ vanish. 
Thus, when $T\rightarrow 0$ and $a\rightarrow\infty$, only 
$G_{{\tt vacuum}}^{wall}({\rm x},{\rm x}')$ is left in eq. (\ref{dnrfeynman}).

Proceeding now as in section \ref{lcircle}, considering $N=2$ and $M\rightarrow 0$, 
eqs.  (\ref{dnrfeynman}) and (\ref{fprime}) are used in eqs. (\ref{edensity}) to (\ref{mfdensity})
to obtain the four components of $\left<T_\mu{}_\nu\right>$, resulting,
\begin{equation}
\left<T_{t}{}_{t}\right>=
\left<T_{t}{}_{t}\right>_{{\tt vacuum}}^{Casimir}+
\left<T_{t}{}_{t}\right>_{{\tt vacuum}}^{wall}+
\left<T_{t}{}_{t}\right>_{{\tt mixed}}^{Casimir}+
\left<T_{t}{}_{t}\right>_{{\tt mixed}}^{wall}+
\left<T_{t}{}_{t}\right>_{{\tt thermal}},
\label{ied}
\end{equation}
and,
\begin{equation}
\left<T_{x}{}_{x}\right>=\left<T_{t}{}_{t}\right>_{{\tt vacuum}}^{Casimir}+
\left<T_{t}{}_{t}\right>_{{\tt mixed}}^{Casimir}+
\left<T_{t}{}_{t}\right>_{{\tt thermal}}, 
\hspace{1.0cm}
\left<T_{t}{}_{x}\right>=\left<T_{x}{}_{t}\right>=0.
\label{icomponents}
\end{equation}
Contribution $\left<T_{t}{}_{t}\right>_{{\tt thermal}}$ is the ``blackbody'' energy density in eq. (\ref{cvted}) and,
\begin{equation}
\left<T_{t}{}_{t}\right>_{{\tt vacuum}}^{Casimir}=-\frac{\pi}{24a^{2}}, 
\hspace{1.0cm}
\left<T_{t}{}_{t}\right>_{{\tt vacuum}}^{wall}=\pm \xi\frac{\pi}{2a^{2}}\csc^{2}\left(\frac{\pi x}{a}\right),
\label{vcomponents}
\end{equation}
are the vacuum energy densities.
Contributions 
$\left<T_{t}{}_{t}\right>_{{\tt mixed}}^{Casimir}$ and 
$\left<T_{t}{}_{t}\right>_{{\tt mixed}}^{wall}$,
that contain double series [see eq. (\ref{feynman7})] and are source of ambiguities,
will be treated shortly.

As already mentioned, this author has not found in the literature
any study of $\left<T_\mu{}_\nu\right>$ for hot scalar radiation in an interval with reflecting edges.
There are though studies of $\left<T_\mu{}_\nu\right>$ at zero temperature \cite{ful07,sah07}, and
the sum of the contributions in eq. (\ref{vcomponents}) is in agreement with 
the vacuum energy density calculated in these references.

The ambiguity in the value of the homogeneous $\left<T_{t}{}_{t}\right>_{{\tt mixed}}^{Casimir}$
corresponds to that  in eqs. (\ref{u}) and (\ref{v}), namely,
\begin{equation}
\left<T_{t}{}_{t}\right>_{{\tt mixed}}^{Casimir}=T^{2}u(2Ta),
\hspace{1.0cm}
\left<T_{t}{}_{t}\right>_{{\tt mixed}}^{Casimir}=T^{2}v(2Ta),
\label{uvcasimir}
\end{equation}
where eq. (\ref{vu}) should be noticed.
The ambiguity in the value of the non homogeneous $\left<T_{t}{}_{t}\right>_{{\tt mixed}}^{wall}$ is new. 
Again, one may sum first over the ``boundary'' number $n$, i.e.,
\begin{equation}
\left<T_{t}{}_{t}\right>_{{\tt mixed}}^{wall}=\mp 2\xi T^{2}\mu(2x/a,Ta),
\hspace{1.0cm}
\mu(p,q):=\frac{2}{\pi}\sum_{m=1}^{\infty}\sum_{n=-\infty}^{\infty}
\frac{m^{2}-q^{2}(p-2n)^{2}}{\left[m^{2}+q^{2}(p-2n)^{2}\right]^{2}}.
\label{mu}
\end{equation}
But also,  one may sum first over the ``thermal'' number $m$, i.e., 
\begin{equation}
\left<T_{t}{}_{t}\right>_{{\tt mixed}}^{wall}=\mp 2\xi T^{2}\nu (2x/a,Ta),
\hspace{1.0cm}
\nu(p,q):=\frac{2}{\pi}\sum_{n=-\infty}^{\infty}\sum_{m=1}^{\infty}
\frac{m^{2}-q^{2}(p-2n)^{2}}{\left[m^{2}+q^{2}(p-2n)^{2}\right]^{2}}.
\label{nu}
\end{equation}
An interesting fact to point out is that for $\xi=0$ (i.e., for minimal and conformal couplings),
the ``wall'' ambiguity in  eqs. (\ref{mu}) and (\ref{nu}) disappears; whereas the ``Casimir'' ambiguity
in eq. (\ref{uvcasimir}) remains. In fact, when $\xi=0$,  the expressions for the components of
$\left<T_\mu{}_\nu\right>$ in an interval with reflecting edges are given by those  for the circle in
eqs. (\ref{components}), (\ref{birrel}) and (\ref{dowker}), after replacing $a$ by $2a$
[see eqs. (\ref{ied}) and (\ref{icomponents})].

Note that the symmetry,
\begin{equation}
\mu(2-p,q)=\mu(p,q), \hspace{1.0cm} \nu(2-p,q)=\nu(p,q),
\label{symmetry}
\end{equation}
was already expected since
the two identical reflecting walls are sitting at $x=0$ and at $x=a$.
By setting $p=2x/a$, then $x=0$  and $x=a$ correspond to $p=0$ and $p=2$, respectively.
Regarding $\left<T_{t}{}_{t}\right>_{{\tt mixed}}^{wall}$ in eqs. (\ref{mu}) and (\ref{nu}),
it follows that one can consider $p$ running from 0 to 1, using then eq. (\ref{symmetry}) 
to determine 
$\left<T_{t}{}_{t}\right>_{{\tt mixed}}^{wall}$ in the other half of the interval,
i.e., for $x>a/2$.
[Clearly the same remark applies to $\left<T_{t}{}_{t}\right>_{{\tt vacuum}}^{wall}$
in eq. (\ref{vcomponents}).]

Comparison of $\mu$ and $\nu$ in eqs. (\ref{mu}) and (\ref{nu})
with  $u$ and $v$ in eqs. (\ref{u}) and (\ref{v}) shows that ($2q$ below is the argument of functions
$u$ and $v$),
\begin{equation}
\mu(0,q)=2u(2q)+\frac{\pi}{3},\hspace{1.0cm}
\nu(0,q)=2v(2q)+\frac{\pi}{3}.
\label{comparison}
\end{equation}
Using now eq. (\ref{vu}), eq. (\ref{comparison}) yields,
\begin{equation}
\nu(0,q)=\mu(0,q)+\frac{1}{q}.
\label{0numu}
\end{equation}
Then, taking into account eq. (\ref{symmetry}), it follows that the relation in eq. (\ref{0numu})
holds also when $p=2$. Indeed, these facts suggest that the relation in eq. (\ref{0numu}) may
hold for arbitrary $p\in [0,2]$, i.e.,
\vspace{0.2cm}

{\bf Conjecture:}
\begin{equation}
\nu(p,q)=\mu(p,q)+\frac{1}{q}.
\label{numu}
\end{equation}
This author does not have a proof of the equality in eq. (\ref{numu}) for arbitrary $p$, 
although there is
strong numerical evidence that supports it \cite{mat17}. The main reason to display the conjecture as
in eq. (\ref{numu}) is to check its consistency with results that will appear
along the text 
\footnote{It should be stressed that there is a proof of eq. (\ref{numu})
when $p=0$ and $p=2$, as has been shown. The proof when $p$ is arbitrary possibly involves
some generalization of ``Schl\"{o}milch's formula'', which may turn out to be a hard task.
}.

The summation over $n$ in eq. (\ref{mu}) can be evaluated (by using, e.g., ref. \cite{mat17}),
yieding, 
\begin{equation}
\mu(p,q)=-\frac{\pi}{4q^{2}}\sum_{k=1}^{\infty}
\left[\csc^{2}\left\{\frac{\pi(pq-ik)}{2q}\right\}+
\csc^{2}\left\{\frac{\pi(pq+ik)}{2q}\right\}
\right].
\label{mu2}
\end{equation}
By using trigonometric and hyperbolic identities, and after some manipulations, 
eq. (\ref{mu2}) can be recast as,
\begin{equation}
\mu(p,q)=\frac{\pi}{q^{2}}\sum_{k=1}^{\infty}
\frac{\cos(p\pi)\cosh\left(k\pi/q\right)-1}
{\left[\cos(p\pi)-\cosh\left(k\pi/q\right)\right]^{2}}.
\label{mu3}
\end{equation}
The first summation in eq. (\ref{nu}) can also be evaluated \cite{mat17},
resulting,
\begin{eqnarray}
&&\nu(p,q)=\frac{\pi}{4q^{2}}\csc^{2}\left(\frac{p\pi}{2}\right)-\pi\,{\rm cosech}^{2}(pq\pi)
\nonumber
\\
&&\hspace{1.3cm}
-\pi\sum_{k=1}^{\infty}\left[{\rm cosech}^{2}\{q\pi (2k-p)\}+{\rm cosech}^{2}\{q\pi (2k+p)\}
\right],
\label{nu2}
\end{eqnarray}
where it should be noticed that although each one of the first two terms diverges as $p\rightarrow 0$,
their sum remains finite, i.e.,
\begin{equation}
\lim_{p\rightarrow 0}
\left[\frac{\pi}{4q^{2}}\csc^{2}\left(\frac{p\pi}{2}\right)-\pi\,{\rm cosech}^{2}(pq\pi)\right]=
\frac{\pi}{12 q^{2}}+\frac{\pi}{3}.
\label{cancelation}
\end{equation}

Taking into account the dependence on $\xi$ in eqs. (\ref{vcomponents}), (\ref{mu}) and  (\ref{nu}), one sees that 
the stationary 
$\left<T_\mu{}_\nu\right>$ 
in eqs. (\ref{ied}) and (\ref{icomponents}) is tracelles when $\xi=0$, and that since the
stress $\left<T_{x}{}_{x}\right>$ is homogeneous,
$\left<T^\mu{}^\nu\right>_{\hspace{-0.05cm},\nu}=0$.

As for the case of the circle in the previous section, the physics of $\left<T_\mu{}_\nu\right>$ 
in eqs. (\ref{ied}) and (\ref{icomponents})
can be better studied by looking at the thermal behaviors of $\left<T_\mu{}_\nu\right>$
corresponding to $Ta\ll 1$ and $Ta\gg1$. However, 
it is worth remarking that
now $\left<T_\mu{}_\nu\right>$ is non homogeneous,
i.e., its value near one of the walls (say, $x\approx 0$) is different from that in the bulk of the box
(say, $x\approx a/2$). 
Before embarking in this study, one has to decide which summations are going to be 
considered first in the expressions of $\left<T_{t}{}_{t}\right>_{{\tt mixed}}^{Casimir}$ and of
$\left<T_{t}{}_{t}\right>_{{\tt mixed}}^{wall}$ [see eqs. (\ref{uvcasimir}), (\ref{mu}) and (\ref{nu})].
For the sake of consistency, the same order of summation will be taken in both expressions.
Note that the expression for $\left<T_{x}{}_{x}\right>$  
can be obtained from that for $\left<T_{t}{}_{t}\right>$ simply by setting $\xi=0$ 
in the latter
[see eq. (\ref{icomponents})], i.e.,
\begin{equation}
\left<T_{x}{}_{x}\right>=\left<T_{t}{}_{t}\right>_{\xi=0},
\label{p}
\end{equation}
showing explicitly the independence of the stress on $\xi$ and thus on the 
type of boundary condition.

\subsubsection{Summing first over n}
\label{nlinterval}

Looking at eq. (\ref{mu3}), it follows quickly that,
\begin{equation}
\mu(p,q\rightarrow 0)=\frac{2\pi}{q^{2}}\cos (p\pi)e^{-\pi/q},
\label{musq}
\end{equation}
where, as already mentioned, exponential smaller terms are being omitted.
Now, using eq. (\ref{usq}) in eq. (\ref{uvcasimir}), and eq. (\ref{musq}) in eq. (\ref{mu}),
at the same time noting eqs. (\ref{cvted}) and (\ref{vcomponents}),
then eq. (\ref{ied}) gives,
\begin{eqnarray}
&&
\hspace{-1.5cm}
\left<T_{t}{}_{t}\right>=-\frac{\pi}{24a^{2}}\pm \xi\frac{\pi}{2a^{2}}\csc^{2}\left(\frac{\pi x}{a}\right)
+\left[1\mp 4\xi\cos\left(\frac{2\pi x}{a}\right)\right]\frac{\pi}{a^{2}}e^{-\pi/Ta},
\hspace{1.0cm} 
Ta\ll 1,
\label{inedsq}
\end{eqnarray}
which holds not only for low temperatures, but also for arbitrary temperatures and small enough boxes.
It should be noted that, although the correction in eq. (\ref{inedsq}) to the vacuum energy density is
exponential small, it will turned out to be relevant when certain thermodynamic issues are addressed
ahead in the text.

Regarding the non homogeneous energy density in eq. (\ref{inedsq}), two places in the box are
of particular interest. Namely, very close to one wall (say, the wall at $x=0$),
\begin{equation}
\left<T_{t}{}_{t}\right>=\pm \frac{\xi}{2\pi x^{2}}+(1\mp 4\xi)
\left[
-\frac{\pi}{24a^{2}}+\frac{\pi}{a^{2}}e^{-\pi/Ta}\right],
\hspace{1.0cm} Ta\ll 1,\hspace{0.5cm} x/a\ll 1,
\label{inedsqsp}
\end{equation}
and at the middle of the box,
\begin{equation}
\left<T_{t}{}_{t}\right>=-\frac{\pi}{24 a^{2}}\pm \xi\frac{\pi}{2a^{2}}
+(1\pm 4\xi)\frac{\pi}{a^{2}}e^{-\pi/Ta},
\hspace{1.8cm} Ta\ll 1,\hspace{0.5cm} x=a/2.
\label{inedsqm}
\end{equation}
When $\xi\neq 0$, the first term in eq. (\ref{inedsqsp}) carries a non integrable divergence,
corresponding to  $x\rightarrow 0$, which is well known in the literature of vacuum fluctuations
in boxes with reflecting walls \cite{ful07}. In the bulk of the box, it is seen from eq. (\ref{inedsqm})
that the ``Casimir'' vacuum  energy density can be substantially modified by a non vanishing $\xi$.

In order to obtain the behavior corresponding to $Ta\gg 1$, one can proceed essentially along the
same steps that led from eq. (\ref{ied}) to eq. (\ref{inedsq}). But now it should be noticed that,
keeping $0<p\lsim 1$, eq. (\ref{nu2}) yields,
\begin{equation}
\nu(p,q\rightarrow\infty)=\frac{\pi}{4q^{2}}\csc^{2}\left(\frac{p\pi}{2}\right)-\pi\,{\rm cosech}^{2}(pq\pi)
-8\pi e^{-4\pi q}\cosh(2pq\pi).
\label{nubq}
\end{equation}
Then, using eq. (\ref{ubq}) in eq. (\ref{uvcasimir}), and eq. (\ref{nubq}) in  eq. (\ref{numu}),
it results that, 
\begin{eqnarray}
&&
\hspace{-1.5cm}
\left<T_{t}{}_{t}\right>=\pm 2\pi\xi T^{2}{\rm cosech}^{2}(2\pi Tx)+\frac{\pi}{6}T^{2}
-(1\mp 4\xi)\frac{T}{2a}
\nonumber
\\
&&\hspace{3.3cm}
-4\pi T^{2}\left[
1\mp 4\xi\cosh(4\pi Tx)
\right]e^{-4\pi Ta},
\hspace{1.0cm} 
Ta\gg 1,
\label{inedbq}
\end{eqnarray}
for $0<x\lsim a/2$. As it stands, eq. (\ref{inedbq}) is also a conjecture, except when $x/a\rightarrow 0$,
in which case eq. (\ref{0numu}) can be used, i.e.,
\begin{equation}
\left<T_{t}{}_{t}\right>=\pm \frac{\xi}{2\pi x^{2}}+(1\mp 4\xi)
\left[
\frac{\pi}{6}T^{2}-\frac{T}{2a}-4\pi T^{2}e^{-4\pi Ta}\right],
\hspace{1.0cm} Ta\gg 1,\hspace{0.5cm} x/a\ll 1.
\label{inedbqsp}
\end{equation}
It is worth noting that the ``Casimir'' vacuum energy density in eq. (\ref{inedsqsp}) 
[first term between right brackets] and
the ``blackbody'' energy density in eq. (\ref{inedbqsp}) play similar roles, with the latter
diminished by a linear term in $T/a$.

The behavior of 
$\left<T_{t}{}_{t}\right>$ in the bulk, for high temperatures or large boxes
[that should be confronted with that in eq. (\ref{inedsqm})] can be obtained from eq. (\ref{inedbq}), resulting,
\begin{equation}
\left<T_{t}{}_{t}\right>=\frac{\pi}{6}T^{2}-
(1\mp 4\xi)\frac{T}{2a}\pm 16\pi\xi T^{2}e^{-2\pi Ta},
\hspace{1.8cm} Ta\gg 1,\hspace{0.5cm} x=a/2,
\label{inedbqm}
\end{equation}
which is essentially ``blackbody'', but corrected by a term linear in $T/a$ that depends on $\xi$.
[Recall  that eq. (\ref{inedbqm}), 
though numerically supported \cite{mat17},
is a conjecture.]

As has been previously mentioned, expressions for the stress $\left<T_{x}{}_{x}\right>$
can be obtained from those for $\left<T_{t}{}_{t}\right>$ above, 
as prescribed in eq. (\ref{p}).
An example is perhaps instructive. For instance, eq. (\ref{inedsqm}) corresponds to,
\begin{equation}
\left<T_{x}{}_{x}\right>=-\frac{\pi}{24 a^{2}}
+\frac{\pi}{a^{2}}e^{-\pi/Ta},
\hspace{1.8cm} Ta\ll 1,\hspace{0.5cm} x=a/2,
\label{instresssqm}
\end{equation}
which is essentially the familiar ``Casimir'' effect: vacuum force attracting  two reflecting walls.

\subsubsection{Summing first over m}
\label{mlinterval}

Starting with eq. (\ref{ied}) and assuming  eq. (\ref{numu}),
it is straightforward to show that
$\left<T_{t}{}_{t}\right>$
corresponding to ``summing first over $m$'' should be obtained from that
for ``summing first over $n$'' by adding  the following  homogeneous term, 
\begin{equation}
(1\mp 4\xi) \frac{T}{2a}.
\label{term}
\end{equation}
According to this prescription, e.g., eqs. (\ref{inedsqsp}) and (\ref{inedsqm}) lead to, 
\begin{equation}
\left<T_{t}{}_{t}\right>=\pm \frac{\xi}{2\pi x^{2}}+(1\mp 4\xi)
\left[
-\frac{\pi}{24a^{2}}+
\frac{T}{2a}+
\frac{\pi}{a^{2}}e^{-\pi/Ta}\right],
\hspace{1.0cm} Ta\ll 1,\hspace{0.5cm} x/a\ll 1,
\label{imedsqsp}
\end{equation}
and,
\begin{equation}
\left<T_{t}{}_{t}\right>=-\frac{\pi}{24 a^{2}}\pm \xi\frac{\pi}{2a^{2}}
+(1\mp 4\xi)\frac{T}{2a}+(1\pm 4\xi)\frac{\pi}{a^{2}}e^{-\pi/Ta},
\hspace{0.7cm} Ta\ll 1,\hspace{0.5cm} x=a/2,
\label{imedsqm}
\end{equation}
respectively. Now, whereas eq. (\ref{imedsqm}) is a conjecture; eq. (\ref{imedsqsp}) is not [see eq. (\ref{0numu})].

In order to address the regime $Ta\gg 1$, instead of considering the prescription associated with eq. (\ref{term}),
one can start again with eq. (\ref{ied}), using eq. (\ref{nubq}) in eq. (\ref{nu}),
to show that $\left<T_{t}{}_{t}\right>$ is given by eq. (\ref{inedbq}) with the term linear in $T/a$ missing.
Therefore, it follows that,
\begin{equation}
\left<T_{t}{}_{t}\right>=\pm \frac{\xi}{2\pi x^{2}}+(1\mp 4\xi)
\left[
\frac{\pi}{6}T^{2}-4\pi T^{2}e^{-4\pi Ta}\right],
\hspace{1.0cm} Ta\gg 1,\hspace{0.5cm} x/a\ll 1,
\label{imedbqsp}
\end{equation}
and that,
\begin{equation}
\left<T_{t}{}_{t}\right>=\frac{\pi}{6}T^{2}
\pm 16\pi\xi T^{2}e^{-2\pi Ta},
\hspace{3.2cm} Ta\gg 1,\hspace{0.5cm} x=a/2.
\label{imedbqm}
\end{equation}
The remark just after eq. (\ref{inedbqsp}) applies here as well. That is,
the ``Casimir'' vacuum energy density in eq. (\ref{imedsqsp}) and
the ``blackbody'' energy density in eq. (\ref{imedbqsp}) play similar roles, 
but now it is the former that is shifted by a linear term in $T/a$.
Recall that $\left<T_{x}{}_{x}\right>$ follows immediately from eq. (\ref{p}).

\subsection{Local thermodynamic stability}
\label{istability}

Using the same set up as that in section \ref{cstability} (i.e., 
a small segment of the reflecting box where the temperature
$T_{in}$, inside, is slightly different from the temperature $T_{out}$, outside),
the following investigation of the regimes $Ta\ll 1$ and $Ta\gg 1$ 
will show that not all values of the coupling parameter $\xi$ are consistent with
local thermodynamic stability. As the stress $\left<T_{x}{}_{x}\right>$ will be needed in 
eq. (\ref{power}), it is worth recalling once more that it can be obtained from the
corresponding $\left<T_{t}{}_{t}\right>$ by simply taking $\xi=0$
[see eq. (\ref{p})].
\subsubsection{Summing first over n}
\label{nistability}
Looking at $\left<T_{t}{}_{t}\right>$ in
eq. (\ref{inedsq}), which holds when $Ta\ll 1$, it follows that  eq. (\ref{power}) yields
$\Phi>0$ if $T_{in}>T_{out}$, i.e., energy leaves the segment. Thus, to ensure
that thermodynamic equilibrium is restored
(in other words, to ensure that $T_{in}$ drops), one must have that 
$\partial _{T}\left<T_{t}{}_{t}\right> >0$ everywhere in the box, then resulting
from eq. (\ref{inedsq}) the constraint,
\begin{equation}
-\frac{1}{4}\leq\xi\leq\frac{1}{4},
\label{bounds1}
\end{equation}
regardless the type of boundary condition (i.e., whether it is Dirichlet's or Neumann's).
It should be remarked that eq. (\ref{bounds1}) would also follow from 
eqs. (\ref{inedsqsp}) and (\ref{inedsqm}).
It should also be pointed out that eq. (\ref{bounds1}) includes the
minimal and conformal couplings, i.e., $\xi=0$; but this is not always the case,
since, for example, in higher number of dimensions 
when a single Dirichlet wall is present
(see ref. \cite{del15}), conformal coupling
is allowed whereas minimal coupling is not.

Dealing with $Ta\gg 1$ now, one takes $\xi=0$ in eq. (\ref{inedbqsp}) to obtain $\left<T_{x}{}_{x}\right>$,
and uses again  eq. (\ref{power}) to conclude that $\Phi>0$ if $T_{in}>T_{out}$.
Thus, requiring that $\partial _{T}\left<T_{t}{}_{t}\right> >0$ in eq. (\ref{inedbqsp}), it results that $\xi$
must be such that,
\begin{equation}
{\tt Neumann:} \; \xi\geq -1/4, 
\hspace{2.0cm}
{\tt Dirichlet:} \; \xi\leq 1/4,
\label{bounds2}
\end{equation}
which are constraints consistent
with eq. (\ref{bounds1}) but less stringent. 
It is worth noting that since 
eq. (\ref{inedbqm}) is essentially ``blackbody'' it does not set any constraint on 
the coupling parameter $\xi$.
\subsubsection{Summing first over m}
\label{mistability}

Taking into account eq. (\ref{imedsqsp}) and
eq. (\ref{power}), it follows that $\Phi>0$ if $T_{in}>T_{out}$. By requiring 
$\partial _{T}\left<T_{t}{}_{t}\right> >0$ in eq. (\ref{imedsqsp}), it leads again to
\footnote{By considering a single reflecting wall at $x=0$,
the constraints in eq. (\ref{bounds2}) are also required \cite{del15,mor19}.}
eq. (\ref{bounds2}),
instead of the more stringent 
constraint in eq. (\ref{bounds1}). 
Clearly the same outcome follows from eq. (\ref{imedsqm}).

Turning now to eq. (\ref{imedbqsp}), an identical analysis yields once more eq. (\ref{bounds2}),
whereas eq. (\ref{imedbqm}) sets no constraint on $\xi$. Note that eq. (\ref{bounds2})
includes the minimal and conformal couplings in the constraints associated with both boundary conditions.

\subsection{Thermodynamics}
\label{gwalls}
Perhaps some of the most interesting aspects of the 
ambiguities addressed in this paper are in the thermodynamics
of the scalar radiation in an interval with reflecting edges.
In order to obtain the internal energy $U$ one still uses eq. (\ref{ienergy}),
but now, unlike the case in section \ref{gcircle}, 
$\left<T_{t}{}_{t}\right>$
in eq. (\ref{ied}) is non homogeneous due to the presence of the terms 
$\left<T_{t}{}_{t}\right>_{{\tt vacuum}}^{wall}$ 
and $\left<T_{t}{}_{t}\right>_{{\tt mixed}}^{wall}$
when $\xi\neq 0$ [note eqs. (\ref{vcomponents}), (\ref{mu}) and (\ref{nu})]. 
In fact, as mentioned previously in the paper [see eq. (\ref{inedsqsp}) and text after it],
$\left<T_{t}{}_{t}\right>_{{\tt vacuum}}^{wall}$ in eq. (\ref{vcomponents})
carries non integrable divergences, 
and its integration from $x=0$ to $x=a$ requires regularization,
which when properly implemented yields a vanishing contribution \cite{ful07}.
Thus, by considering eq. (\ref{ied}) in eq. (\ref{ienergy}), the only non trivial
integration is,
\begin{equation}
\int_{0}^{a}\left<T_{t}{}_{t}\right>_{{\tt mixed}}^{wall}dx=
\mp\xi a T^{2}\int_{0}^{2}\mu(p,Ta)dp,
\hspace{1.0cm}
\int_{0}^{a}\left<T_{t}{}_{t}\right>_{{\tt mixed}}^{wall}dx=
\mp\xi a T^{2}\int_{0}^{2}\nu(p,Ta)dp,
\label{mixedie}
\end{equation}
corresponding to the ambiguity in eqs. (\ref{mu}) and (\ref{nu}).

Regarding the integration of $\mu$ over $p$ in eq. (\ref{mixedie}), by noticing that,
$$
\int
\frac{\cos(p\pi)\cosh\left(k\pi/q\right)-1}
{\left[\cos(p\pi)-\cosh\left(k\pi/q\right)\right]^{2}} dp=
\frac{1}{\pi}\frac{\sin(p\pi)}{\cosh\left(k\pi/q\right)-\cos(p\pi)},
$$
it follows from eq. (\ref{mu3}) that,
\begin{equation}
\int_{0}^{2}\mu(p,q)dp=0,
\label{itmu}
\end{equation}
and therefore the first integration in eq. (\ref{mixedie}) gives
a vanishing contribution to the internal energy $U$. It is worth remarking that a check of consistence
can be implemented by considering eq. (\ref{musq}) which  leads promptly to 
eq. (\ref{itmu}), as it should.

Regarding now the integration of $\nu$ over $p$ in eq. (\ref{mixedie}), one uses eq. (\ref{symmetry})
to write,
\begin{equation}
\int_{0}^{2}\nu(p,q)dp=2\int_{0}^{1}\nu(p,q)dp.
\label{itnu1}
\end{equation}
Looking at the expression for $\nu$ in eq. (\ref{nu2}), it is a simple matter to show
that,
\begin{equation}
\int_{0}^{1}
\left[
\frac{\pi}{4q^{2}}\csc^{2}\left(\frac{p\pi}{2}\right)-\pi\,{\rm cosech}^{2}(pq\pi)
\right] dp=\frac{1}{q}\coth(q\pi),
\label{itnu2}
\end{equation}
and that,
\begin{equation}
\int_{0}^{1}
{\rm cosech}^{2}\{q\pi(2k\pm p)\} dp=
\mp\frac{1}{q\pi}\left[\coth\{q\pi (2k\pm 1)\}-\coth\{q\pi 2k\}
\right].
\label{itnu3}
\end{equation}
Then one uses eq. (\ref{itnu3}) to integrate the series in eq. (\ref{nu2}), i.e.,
\begin{eqnarray}
&&
\hspace{-2.0cm}
-\pi\sum_{k=1}^{\infty}
\int_{0}^{1}
\left[{\rm cosech}^{2}\{q\pi (2k-p)\}+{\rm cosech}^{2}\{q\pi (2k+p)\}
\right]dp=
\nonumber
\\
&&
-\frac{1}{q}\sum_{k=1}^{\infty}
\left[
\coth\{q\pi (2k-1)\}-\coth\{q\pi (2k+1)\}
\right]=
\frac{1}{q}-\frac{1}{q}\coth(q\pi).
\label{itnu4}
\end{eqnarray}
To complete, by adding eqs. (\ref{itnu2}) and (\ref{itnu4}), eq. (\ref{itnu1}) yields,
\begin{equation}
\int_{0}^{2}\nu(p,q)dp=\frac{2}{q},
\label{itnu5}
\end{equation}
which should be compared with eq. (\ref{itmu}).
A check of consistency can be done using eq. (\ref{nubq}),
by integrating $q\nu$ over $p$,
and then taking $q\rightarrow\infty$, resulting 2 as in eq. (\ref{itnu5}).
It should be also noted that eqs. (\ref{itmu}) and (\ref{itnu5}) offer an opportunity
to check the consistency of the conjecture in eq. (\ref{numu}). Namely, by integrating
both sides of eq. (\ref{numu}), with no surprises.

Finally, one uses eqs. (\ref{itmu}) and (\ref{itnu5}) in the expressions in eq. (\ref{mixedie}), to obtain the
following contrasting contributions,
\begin{equation}
\int_{0}^{a}\left<T_{t}{}_{t}\right>_{{\tt mixed}}^{wall}dx=0,
\hspace{1.0cm}
\int_{0}^{a}\left<T_{t}{}_{t}\right>_{{\tt mixed}}^{wall}dx=
\mp 2\xi T ,
\label{mixedie2}
\end{equation}
respectively. 

In order to obtain $U$ in eq. (\ref{ienergy}), one goes back to eq. (\ref{ied}),
recalling that $\left<T_{t}{}_{t}\right>_{{\tt vacuum}}^{wall}$ 
does not contribute with $U$. Then, by taking into account eqs. (\ref{cvted}), (\ref{vcomponents}), 
(\ref{uvcasimir}) and (\ref{mixedie2}), it results $U$ in eq. (\ref{ie-rwall})
(corresponding to ``summing first over $n$''), or $U$ in eq. (\ref{ie-rwall-zeromode})
(corresponding to ``summing first over $m$''). Note that eqs. (\ref{ie-rwall}) and (\ref{ie-rwall-zeromode})
differ from each other by the term,
\begin{equation}
(1\mp 4\xi)\frac{T}{2},
\label{term2}
\end{equation}
which is consistent with the difference of the corresponding
energy densities in eq. (\ref{term}).

Certain features of the ambiguity in eqs. (\ref{ie-rwall}) and (\ref{ie-rwall-zeromode})
were addressed in section \ref{introduction}. For example, it was mentioned that 
eq. (\ref{ie-rwall}) is the result found in the literature \cite{amb83,lim09},
and that eq. (\ref{ie-rwall-zeromode}) violates the classical result that 
the internal energy $U$ 
should not depend on $\xi$ in the background considered here \cite{ful07}.
Proceeding as in section \ref{gcircle}, the following analysis concerns
thermodynamic aspects of this ambiguity when $Ta\ll 1$, and $Ta\gg 1$.
\subsubsection{Summing first over n}
\label{ngwalls}
It can be quickly checked that eq. (\ref{ie-rwall}) is obtained
from eq. (\ref{ie-circle}), which applies to the circle,
by replacing $a$ by $2a$ and halving the final expression.
In fact, the quantities $U$, $F$ and $S$ here can be obtained from those
in section \ref{ngcircle} using this prescription. It results then that
the thermodynamics of the scalar radiation in an interval with reflecting edges,
according to ``summing first over $n$'', is the same as that discussed in section \ref{ngcircle}.
In particular, the third law of thermodynamics is also satisfied.

By integrating  $\left<T^{\mu}{}_{\mu}\right>$ over the box
[note eqs. (\ref{ied}) and (\ref{icomponents})], due to the first integration
in eq. (\ref{mixedie2}), it results that,
$U-\left<T_{x}{}_{x}\right>a=0$.
As $\left<T_{x}{}_{x}\right>$ is the thermodynamic pressure ${\tt p}$, the equation 
of state in eq. (\ref{estate}) holds here as well.
\subsubsection{Summing first over m}
\label{mgwalls}
When $\xi=0$ (minimal and conformal couplings),
it is seen that eq. (\ref{ie-rwall-zeromode}) is obtained
from eq. (\ref{ie-zeromode}), also
by replacing $a$ by $2a$ and halving the final expression. Quantities $U$,
$F$ and $S$ result from the corresponding quantities
in section \ref{mgcircle} by using this same prescription. Then, when $\xi=0$, 
thermodynamics of the scalar radiation in an interval with reflecting edges,
according to ``summing first over $m$'', resembles very much that in section \ref{mgcircle},
including violation of the third law.

However, when $\xi\neq 0$, new issues come about.
As mentioned above, one can obtain $U$ here by adding the term in eq. (\ref{term2})
to the corresponding expression for $U$ in section {\ref{ngwalls}}. Beginning with the regime
$Ta\ll 1$, it results that,
\begin{equation}
U(T,a)=-\frac{\pi}{24a}+
(1\mp 4\xi)\frac{T}{2}
+\frac{\pi}{a}e^{-\pi/Ta},
\hspace{1.0cm} 
Ta\ll 1.
\label{mesq-walls}
\end{equation}
The internal energy in eq. (\ref{mesq-walls}), which holds for low temperatures or small boxes, offers
a good opportunity to confront the local thermodynamic stability in section \ref{mistability}
with the global one, as remarked in section \ref{introduction}. Since global thermodynamic stability
requires a positive heat capacity, i.e., $C=(1\mp 4\xi)/2+\cdots>0$, one ends up consistently with the bounds on $\xi$ in eq. (\ref{bounds2}) [see text in section \ref{mistability}].

At this point, it is worth noting that
by integrating  $\left<T^{\mu}{}_{\mu}\right>$ over the box again, but now using the second integration
in eq. (\ref{mixedie2}), it results the following equation of state,
\begin{equation}
U-{\tt p}a=\mp 2\xi T,
\label{estate2}
\end{equation}
instead of that in eq. (\ref{estate}). 
It should be remarked that ${\tt p}$ in eq. (\ref{p}) does not
depend on $\xi$.

Considering now eqs. (\ref{mesq-walls}) and (\ref{estate2}), eq. (\ref{fu})
is integrated, introducing a positive length scale $\ell$ that must not depend either on  $T$ or $a$, i.e.,
\begin{equation}
F(T,a)=-\frac{\pi}{24a}-(1\mp 4\xi)\frac{T}{2}\ln(2T\ell)-\frac{T}{2}\ln \frac{a}{\ell}
-Te^{-\pi/Ta},
\hspace{1.0cm} 
Ta\ll 1.
\label{mfsqwalls}
\end{equation}
It should be pointed out that $\ell$ arises only if $\xi\neq 0$. 
The entropy corresponding to eq. (\ref{mfsqwalls}) is given by,
\begin{equation}
S(T,a)=\ln \sqrt{\frac{a}{\ell}}
+\frac{1}{2}(1\mp 4\xi)\left[\ln(2T\ell)+1\right]+\left(\frac{\pi}{Ta}+1\right)e^{-\pi/Ta},
\hspace{1.0cm} 
Ta\ll 1,
\label{mssqwalls}
\end{equation}
which remains finite as $T\rightarrow 0$ for $\xi =1/4$ in the case of Dirichlet's 
boundary condition, and for $\xi= -1/4$ in the case of  Neumann's
boundary condition, i.e., 
$S(T\rightarrow 0,a)=\ln \sqrt{a/\ell}$.
Since $a/\ell$ is not a ``universal constant'',
the third law of thermodynamics is still violated
\footnote{It is worth remarking that according to some authors
this is not a fault, though \cite{wal97}.}.

Considering now the regime $Ta\gg 1$, by adding eq. (\ref{term2})
to the corresponding expression for $U$ in section {\ref{ngwalls}},
one ends up with,
\begin{equation}
U(T,a)=\frac{\pi}{6}aT^{2}\mp2\xi T-4\pi aT^{2}e^{-4\pi Ta},
\hspace{1.0cm} 
Ta\gg 1.
\label{mebq-walls}
\end{equation}
At this point, a remark that makes connection with an earlier paper is in order.
By working with a single reflecting wall in ref. \cite{mor19}, this author arrived to 
$U=\pi a T^{2}/6\mp \xi T$ which, when compared with eq. (\ref{mebq-walls}), it
suggests that the factor of two in $\mp2\xi T$ is due to the presence of a second wall
(as was conjectured in ref. \cite{mor19}). Note that in ref. \cite{mor19} only 
the summation over the ``thermal'' number $m$ appears and thus the ambiguity 
``summing first over $n$'' versus ``summing first over $m$''
is not apparent.

By taking into account eq. (\ref{estate2}), one sees that 
${\tt p}$ is given by dropping the term $\mp2\xi T$ in eq. (\ref{mebq-walls})
and dividing the resulting expression by $a$. The corresponding free energy is given by,
\begin{equation}
F(T,a)=-\frac{\pi}{6}aT^{2}\pm2\xi T\ln (2T\ell)
-Te^{-4\pi Ta},
\hspace{1.0cm} 
Ta\gg 1,
\label{mfbq-walls}
\end{equation}
where the length scale $\ell$ arises again [see eq. (\ref{mfsqwalls})].
To complete,
the behavior of the entropy at high temperatures or for 
large boxes follows from eq. (\ref{mfbq-walls}), namely,
\begin{equation}
S(T,a)=\frac{\pi}{3}aT\mp 2\xi\left[\ln(2T\ell)+1\right]
-(4\pi Ta-1)e^{-4\pi Ta},
\hspace{1.0cm} 
Ta\gg 1.
\label{msbq-walls}
\end{equation}

\section{Further discussion}
\label{conclusion}

This paper investigated the finite temperature 
$\left<T_\mu{}_\nu\right>$ of a massless scalar field on a
circle and in an interval with reflecting edges
(with Dirichlet's or Neumann's boundary conditions at the edges).
In so doing, it was shown that  
$\left<T_\mu{}_\nu\right>$ involves double series which,
due to the number of dimensions of the spacetime under consideration (i.e., $N=2$), are not absolutely convergent
and that this fact is connected with ambiguities in
the calculation of $\left<T_\mu{}_\nu\right>$. Namely, the order
in which the two summations are evaluated leads to different results:
summing first over the ``boundary'' number $n$, versus summing first over the ``thermal'' number $m$.
By studying the associated thermodynamics of each contrasting expressions for $\left<T_\mu{}_\nu\right>$, it was found that in the case of the circle the ambiguity corresponds to the classic debate in the literature of whether or not zero modes should be ignored in the computation of partition functions. In the case of the interval with reflecting edges, 
one of the (non homogeneous) contrasting expressions for
$\left<T_\mu{}_\nu\right>$ leads to the thermodynamics reported in the literature (obtained by using the partition function) whose internal energy $U$ does not depend on the curvature coupling parameter $\xi$ (as one would expect from a classical calculation), whereas the other expression for 
$\left<T_\mu{}_\nu\right>$ leads to $U$ that does depend on $\xi$,
which is rather unexpected.

It was shown that the ambiguities reported in this paper are 
nicely connected with classic results on
infinite series which go back to the works of  Ramanujan.
In this context a conjecture was presented whose consistency 
was checked in various instances.

Regarding the asymptotic regimes $Ta\ll 1$ and $Ta\gg 1$,
although the ambiguities
only affect subleading contributions in the contrasting expressions for $\left<T_\mu{}_\nu\right>$,
their thermodynamic consequences are substantial. For example, 
``summing first over $m$'' leads to violation of the third law 
of thermodynamics and, in the case of the interval with reflecting 
edges, also to an internal energy $U$ that depends on $\xi$, as just mentioned.
It should be recalled that 
``summing first over $n$'' is not free of issues either.
In the case of the circle,
it leads to $U$  
that spoils the derivation of the Cardy-Verlinde formula
(see section \ref{introduction}).

In the case of the interval with reflecting edges,
the requirement of local thermodynamic stability led
to different ranges of permissible values for $\xi$
corresponding to ``summing first over $n$''
(where the constraint is more stringent and it is the same for Dirichlet's and Neumann's boundary conditions) and to 
``summing first over $m$''
(where the constraint is less stringent and which depends on the type of reflecting boundary condition). The side 
``summing first over $m$'' of the ambiguity allowed
to confront local and global thermodynamic stability, and consistency
of the constraints over the values of $\xi$ was verified.

Before closing, it is pertinent to rise an issue which may have
already come to mind. As is typical of series that are not absolutely 
convergent, each way of summing the series may lead to different results and,
consequently, to different physics. In the light of this argument
one might wonder the relevance of the particular ways of
evaluating the summations discussed in this paper, namely,
``summing first over $n$'' vs ``summing first over $m$''.
Whereas indeed other ways of evaluating the summations can lead
to new thermodynamics (whose features may be interesting), those considered in this paper are closely connected with matters that have been addressed previously
in the literature, in various contexts, as shown along the text.

This paper followed a line of investigation that has been established
long ago by Brown, Maclay, Dowker and others, which consists in 
``deriving'' blackbody thermodynamics from $\left<T_\mu{}_\nu\right>$.
This author intends to pursue further studies along this philosophy,
aiming to address classic issues that appear when event horizons
are present in the background.

\appendix
\section{Thermal Green functions}
\label{green-functions}

Consider a cavity in an $N$-dimensional flat spacetime, 
\begin{equation}
ds^2=dt^2-dx^2-dy^2-dz^2-\cdots.
\label{metric1}
\end{equation}
One of the walls of the cavity coincides with the plane $x=0$ and another with $x=a>0$.
The other walls, when they exist, are at infinity.
A neutral scalar field $\phi$ with mass $M$ is in the cavity at 
temperature $T$. Then the coordinate $x_{0}:=it$ is taken to be real
with period $\beta=1/T$ (see, e.g., ref. \cite{ful87}). Considering further
$x_{1}:=x$,  $x_{2}:=y$, $x_{3}:=z$, eq. (\ref{metric1}) becomes,
\begin{equation}
ds^2=-dx_{0}^2-dx_{1}^2-dx_{2}^2-dx_{3}^2-\cdots-dx_{N-1}^2,
\label{metric2}
\end{equation}
and the boundary condition,
\begin{equation}
\phi(x_0,x_1,x_2,\cdots,x_{N-1})
=\phi(x_0+\beta,x_1,x_2,\cdots,x_{N-1}),
\label{bc1}
\end{equation}
must be observed.
The (Euclidean) Green function satisfies \cite{dav82},
\begin{equation}
\left(\Box_{{\rm x}}+M^{2}\right)G_{E}({\rm x},{\rm x}')=
\delta\left({\rm x}-{\rm x}'\right),
\label{fe0}
\end{equation}
where
$\Box_{{\rm x}}:=-\partial^2_0-\partial^2_1-\partial^2_2-\partial^2_3-
\cdots-\partial^2_{N-1}$.

\subsection{Periodic boundary condition}
\label{acircle}
The eigenfunctions of 
$\Box_{{\rm x}}+M^{2}$ 
that satisfy eq. (\ref{bc1}) and
the periodic boundary condition, i.e., 
\begin{equation}
\psi(x_0,x_1,x_2,\cdots,x_{N-1})
=\psi(x_0,x_1+a,x_2,\cdots,x_{N-1}),
\nonumber
\end{equation}
are given by,
\begin{equation}
\psi_ {k}({\rm x})=\eta \exp [i(k_{0}x_{0}+k_{1}x_{1}+\cdots+k_{N-1}x_{N-1})],
\label{ef1}
\end{equation}
where $\eta$, $k_{0}=2\pi m/\beta$, $k_{1}=2\pi n/a$, $\cdots$, $k_{N-1}$
are constants, with $m$ and $n$ integers. The corresponding eigenvalues are,
\begin{equation}
E_{k}=k_{0}^{2}+k_{1}^{2}+\cdots +k_{N-1}^{2}+M^{2}.
\label{ev}
\end{equation}
The constant $\eta$ in eq. (\ref{ef1}) is set such that the
Green function in eq. (\ref{fe0}) is given by,
\footnote{The expression in eq. (\ref{eu})
is known as Schwinger's ``proper time'' representation of the finite temperature Green function (see, e.g., ref. \cite{deu79}).} 
\begin{equation}
G_{E}({\rm x},{\rm x}')=
i\sum_{m=-\infty}^{\infty}
\sum_{n=-\infty}^{\infty}
\int_{0}^{\infty}d\tau
\int_{-\infty}^{\infty}dk_{2}\ \cdots
\int_{-\infty}^{\infty}dk_{N-1}
e^{-i\tau E_{k}}\psi_{k}({\rm x})
\psi^{*}_{k}({\rm x}'),
\label{eu}
\end{equation}
and $M^{2}$ is taken to have an infinitesimal imaginary part to
make the integration over $\tau$ in eq. (\ref{eu}) to converge \cite{dew65}.
Then, it follows that,
\begin{eqnarray}
&&\left(\Box_{{\rm x}}+M^{2}\right)
G_{E}({\rm x},{\rm x}')=
-\sum_{m=-\infty}^{\infty}
\sum_{n=-\infty}^{\infty}
\int_{-\infty}^{\infty}dk_{2}\ \cdots
\int_{-\infty}^{\infty}dk_{N-1}
\psi_{k}({\rm x})
\psi^{*}_{k}({\rm x}')
\nonumber
\\
&&
\hspace{4.0cm}\times\int_{0}^{\infty}d\tau\frac{d}{d\tau}
e^{-i\tau E_{k}},
\label{propertime}
\end{eqnarray}
where the integration over $\tau$ yields simply minus unity.
Now, by choosing 
\begin{equation}
|\eta|^{2}=\frac{(2\pi)^{2-N}}{\beta a},
\label{c1}
\end{equation}
and recalling the usual representation of the $\delta$-function, as well as 
Poisson's formula,
\begin{equation}
\sum_{l=-\infty}^{\infty}\delta(\lambda-2\pi l)=
\frac{1}{2\pi}\sum_{l=-\infty}^{\infty} e^{-il\lambda},
\label{poisson}
\end{equation}
it results that the right hand side of eq. (\ref{propertime}) is indeed 
$\delta({\rm x}-{\rm x}')$, as eq. (\ref{fe0}) requires.

A more workable expression for 
$G_{E}({\rm x},{\rm x}')$
in eq. (\ref{eu}) can be obtained.
Noting eq. (\ref{ef1}), two factors arise in  eq. (\ref{eu})
which can be conveniently manipulated as below,
\begin{eqnarray}
\sum_{l=-\infty}^{\infty}
e^{-i\tau (4\pi^2 l^2/p^2)+i(2\pi l/p)\Delta}&=&
\sum_{l=-\infty}^{\infty}\int_{-\infty}^{\infty}d\lambda\
\delta(\lambda-2\pi l)\
e^{-i\tau (\lambda^2/p^2)+i(\lambda/p)\Delta}
\nonumber
\\
&=&
\frac{1}{2\pi}\sum_{l=-\infty}^{\infty}\int_{-\infty}^{\infty}d\lambda
e^{-i\tau (\lambda^2/p^2)+i(\lambda/p)(\Delta-lp)},
\label{factor}
\end{eqnarray}
where eq. (\ref{poisson}) has been used in the last step.
All the integrations can now be performed \cite {gra07}, leading to,
\begin{eqnarray}
&&G_{E}({\rm x},{\rm x}')=
\frac{1}{(2\pi)^{N/2}}M^{\frac{N-2}{2}}
\sum_{m=-\infty}^{\infty}
\sum_{n=-\infty}^{\infty}
(-\sigma^{(m,n)})^{\frac{2-N}{4}}
K_{\frac{N-2}{2}}\left(M\sqrt{-\sigma^{(m,n)}}\right),
\label{green1}
\end{eqnarray}
where [noting the coordinates in eq. (\ref{metric1})],
\begin{equation}
\sigma^{(m,n)}:=(t-t'-im\beta)^2-(x-x'-na)^2-(y-y')^2-(z-z')^2-\cdots,
\label{cvariable}
\end{equation}
and 
$K_{\nu}({\rm z})$ is the
modified Bessel function of the second kind.

Before addressing Dirichlet and Neumann boundary conditions,
it should be mentioned that the Green function in eq. (\ref{green1})
is closed related with the thermal Hadamard function in eq. (2.26)
of ref. \cite{mel13}, where certain aspects of a charged scalar field
are investigated in a background with an arbitrary number of compact dimensions.

\subsection{Dirichlet and Neumann boundary conditions}
\label{arwalls}
Considering now the Dirichlet boundary condition, i.e.,
\begin{equation}
\psi(x_0,x_1=0,x_2,\cdots,x_{N-1})
=\psi(x_0,x_1=a,x_2,\cdots,x_{N-1})=0,
\nonumber
\end{equation}
the eigenfunctions of 
$\Box_{{\rm x}}+M^{2}$, 
that also satisfy eq. (\ref{bc1}), 
are now given by,
\begin{equation}
\psi_ {k}({\rm x})=\eta\sin(k_{1}x_{1}) \exp [i(k_{0}x_{0}+k_{2}x_{2}+\cdots+k_{N-1}x_{N-1})],
\label{ef2}
\end{equation}
whose eigenvalues are those in eq. (\ref{ev}), 
where $k_{0}=2\pi m/\beta$ and  $k_{1}=n\pi/a$, with $m$ and $n$ integers as before.
By using eq. (\ref{ef2}) in  eq. (\ref{eu}), 
noting eq. (\ref{c1}), eq. (\ref{poisson}), the usual representation of the $\delta$-function and the Fourier sine series \cite{arf85},
\begin{equation}
\delta(x-x')=\frac{1}{a}\sum_{n=-\infty}^{\infty}
\sin(n\pi x/a)\sin(n\pi x'/a),
\label{fsine}
\end{equation}
it follows that eq. (\ref{propertime}) becomes eq. (\ref{fe0}) as it should.
One now expands the sine functions in exponentials and manipulates the sums in eq. (\ref{eu})
as in eq. (\ref{factor}). The last step is to evaluate the integrations \cite{gra07}, resulting in,
\begin{eqnarray}
&&\hspace{1.0cm} G_{E}({\rm x},{\rm x}')=
\frac{1}{(2\pi)^{N/2}}M^{\frac{N-2}{2}}
\sum_{m=-\infty}^{\infty}
\sum_{n=-\infty}^{\infty}
\nonumber
\\
&&
\left[(-\sigma_{-}^{(m,n)})^{\frac{2-N}{4}}
K_{\frac{N-2}{2}}\left(M\sqrt{-\sigma_{-}^{(m,n)}}\right)
-(-\sigma_{+}^{(m,n)})^{\frac{2-N}{4}}
K_{\frac{N-2}{2}}\left(M\sqrt{-\sigma_{+}^{(m,n)}}\right)\right],
\label{green2}
\end{eqnarray}
where,
\begin{equation}
\sigma_{\pm}^{(m,n)}:=(t-t'-im\beta)^2-(x\pm x'-2na)^2-(y-y')^2-(z-z')^2-\cdots.
\label{dncvariable}
\end{equation}
At this point, it should be remarked that the term $n=0$ in eq. (\ref{green2})
reproduces consistently the corresponding Green function in ref. \cite{mor15}
(i.e., as $a\rightarrow\infty$). 
By setting $M\rightarrow 0$ and $N=4$ in  eq. (\ref{green2}), 
the Green function in ref. \cite{tad86} is also consistently reproduced.
\footnote{By setting $M\rightarrow 0$ and $N=4$ in  eq. (\ref{green2}) it should also
match the results in ref. \cite{dav82}; however, a typo has been detected 
in eq. (4.38) of ref. \cite{dav82}: ``$an$'' should  be replaced by ``$2an$''.}

Turning now to the Neumann boundary condition, i.e.,
\begin{equation}
\frac{\partial}{\partial x_{1}}\psi(x_0,x_1=0,x_2,\cdots,x_{N-1})
=\frac{\partial}{\partial x_{1}}\psi(x_0,x_1=a,x_2,\cdots,x_{N-1})=0,
\nonumber
\end{equation}
one proceeds as in Dirichlet's above, but now replacing the sine functions in
eqs. (\ref{ef2}) and (\ref{fsine}) by cosine functions. It results a Green function still given
by eq. (\ref{green2}); but with the minus sign between the terms containing Bessel functions
replaced by a plus sign.


\vspace{1cm}
\noindent{\bf Acknowledgements} -- 
This author wishes to thank 
Lucas dos Santos, 
Luis Fernando Mello, 
Marcia Kashimoto, and
Claudemir de Oliveira
for helpful conversations on the convergence 
of double series.
Work partially supported by
``Funda\c{c}\~{a}o de Amparo \`{a} Pesquisa do Estado de Minas Gerais'' (FAPEMIG)
and by ``Coordena\c{c}\~{a}o de Aperfei\c{c}oamento de Pessoal de N\'{\i}vel Superior'' (CAPES).

.



\end{document}